\documentclass[%
 reprint,
superscriptaddress,
 amsmath,amssymb,
 prl,
]{revtex4-2}

\usepackage{comment}
\usepackage{graphicx}
\usepackage{dcolumn}
\usepackage{bm}
\usepackage{xcolor}
\usepackage{hyperref}
\usepackage{multirow}

\newcommand {\FIG}[1]{Fig. \ref{#1}}
\newcommand {\FIGS}[1]{Figs. \ref{#1}}
\newcommand {\EQ}[1]{Eq. (\ref{#1})}

\begin{document}

\title{Shortest-Path Percolation on Random Networks}

\author{Minsuk Kim}
\affiliation{Center for Complex Networks and Systems Research, Luddy School of Informatics, Computing, and Engineering, Indiana University, Bloomington, Indiana, 47408, USA}
\author{Filippo Radicchi}
 \email{filiradi@indiana.edu}
\affiliation{Center for Complex Networks and Systems Research, Luddy School of Informatics, Computing, and Engineering, Indiana University, Bloomington, Indiana, 47408, USA}

\begin{abstract}
We propose a bond-percolation model intended to describe the consumption, and eventual exhaustion, of resources in 
transport networks. Edges forming minimum-length paths connecting demanded origin-destination nodes are removed if below a certain budget. As pairs of nodes are demanded and edges are removed, the macroscopic connected component of the graph disappears, i.e., the graph undergoes a percolation transition. Here, we study such a shortest-path-percolation transition in homogeneous random graphs where pairs of demanded origin-destination nodes are randomly generated, and fully characterize it by means of finite-size scaling analysis. If budget is finite, the transition is identical to the one of ordinary percolation, where a single giant cluster shrinks as edges are removed from the graph; for infinite budget, the transition becomes more abrupt than the one of ordinary percolation, being characterized by the sudden fragmentation of the giant connected component into a multitude of clusters of similar size. 
\end{abstract}

\maketitle

Percolation theory studies the relation between the macroscopic connectedness of a system and its microscopic structure. 
Percolation models are fruitfully applied to many physical systems, e.g., gelation of molecules, diffusion in porous media, and forest fires~\cite{stauffer2018introduction}.
In network science, percolation 
models are traditionally used
to characterize the robustness of 
social, biological, and economic networks~\cite{callaway2000network, cohen2000resilience, albert2000error, dunne2002network, haldane2011systemic, li2021percolation, schroder2018hysteretic}.
The existence of a macroscopic connected component in a network is interpreted as a proxy of its overall function.
The connectedness of the network may be compromised by the deletion or failure of its microscopic components, either nodes (site percolation) or edges (bond percolation). The protocol used to delete the network's microscopic elements defines the specific percolation model at hand.
In the ordinary percolation model, deleted microscopic elements are chosen uniformly at random~\cite{stauffer2018introduction}. Other well-known percolation models include targeted attacks~\cite{albert2000error}, $k$-core percolation~\cite{dorogovtsev2006k}, cascading failures~\cite{buldyrev2010catastrophic}, continuous percolation with discontinuities~\cite{nagler2012continuous}, explosive percolation~\cite{achlioptas2009explosive}, fractional percolation~\cite{schroder2013crackling}, and optimal percolation~\cite{artime2024robustness}.
These protocols can be extended to account for
multiplexity~\cite{buldyrev2010catastrophic, radicchi2015percolation} and higher-order interactions~\cite{sun2023dynamic}.

Percolation-based approaches are popular also in the analysis of  dynamical processes occurring on infrastructural networks, e.g., car congestion in road networks~\cite{li2015percolation, zeng2019switch, zeng2020multiple, cogoni2021stability, hamedmoghadam2021percolation}. By assigning a quality score to each edge (i.e., road segment) and removing edges with quality below a given threshold, the above-mentioned studies focus on how the emergence of congested clusters affects the overall function of a road network. Similar approaches are used to study road networks subject to flooding~\cite{yadav2020resilience, dong2022modest} and sidewalk networks in cities during the pandemic~\cite{rhoads2021sustainable}.

Here, we introduce a bond-percolation model specifically devised to mimic the utilization and progressive exhaustion of a transport network's resources. We named it as the shortest-path-percolation (SPP) model because edges are removed from the network whenever they form paths of minimum length connecting pairs of nodes. A real system that could be described by the SPP model is an airline network where travelers select minimum-cost itineraries connecting their desired origin-destination airports~\cite{zanin2013modelling, verma2016emergence}.

\begin{figure}[!htb]
    \centering
    \includegraphics[width=\columnwidth]{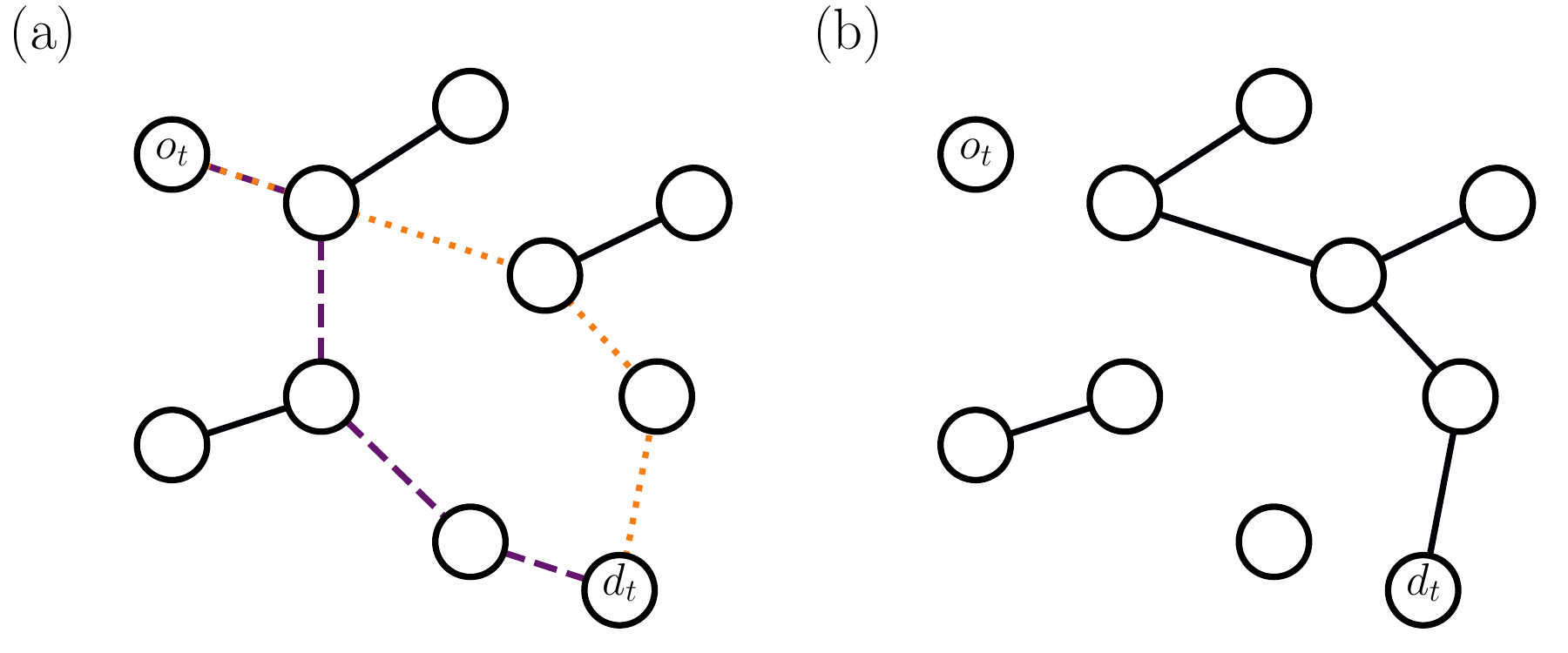}
    \caption{The shortest-path-percolation model. (a) There are two possible shortest paths connecting the origin-destination pair $o_t \to d_t$ demanded by the agent $t$, denoted by orange dotted edges and purple dashed edges, respectively. The length of such shortest paths is $Q_t = 4$. 
    (b) If the maximum length allowed in the SPP model is $C \geq 4$, one of the two shortest paths is selected at random, then all of its edges are removed from the graph. Here, all purple dashed edges are deleted, and the graph fragments into four clusters.}
    \label{figure1}
\end{figure}

The SPP model is defined as follows.
For $t>0$,  we denote with $\mathcal{G}_{t} = (\mathcal{V}, \mathcal{E}_{t})$, composed of $N = |\mathcal{V}|$ nodes and $E_t= |\mathcal{E}_t|$ edges, 
the undirected and unweighted graph available to the agent $t$, and with $o_t \to d_t$ the origin-destination pair demanded by the agent $t$. If at least a path between $o_t$ and $d_t$ exists in $\mathcal{G}_{t}$, we denote with $Q_t$ the length of the shortest one(s). The demand of the agent $t$ can be supplied only if $d_t$ is reachable from $o_t$ and $Q_t \leq C$, where $C > 0$ is a tunable parameter of the model. If one shortest path satisfying this condition is identified (one path is selected at random if more than one exists), namely $(o_t = i_1 , i_2 , \ldots , i_{Q_{t}+1} = d_t)$, all edges in the path are removed from $\mathcal{G}_{t}$, i.e., $\mathcal{E}_{t}  \mapsto \mathcal{E}_{t} \setminus \bigcup_{q=1}^{Q_{t}} (i_{q},  i_{q+1})$, see~\FIG{figure1}. If no path exists between $o_t$ and $d_t$ or if $Q_t > C$, no edge is  removed from the graph. In either case, we copy the graph $\mathcal{G}_{t+1} \mapsto \mathcal{G}_{t}$ and then increase $t \mapsto t + 1$. The process is repeated until no more demand is requested or can be supplied.

The behavior of the SPP model depends on the structure of the graph $\mathcal{G}_1$ and the demand of the agents. Here, for simplicity, we assume that the graph $\mathcal{G}_1$ is an instance of the Erd\H{o}s-R\'{e}nyi (ER) model with exactly $E_1 = \bar{k} N / 2$ edges, with  $\bar{k}$ average degree of the graph. We further assume that the origin-destination nodes $o_t \to d_t$ demanded by the agent $t$ are chosen uniformly at random. 
These assumptions are not reasonable for the study of a real infrastructure and are made with the sole purpose of understanding the physics of the SPP model. They in fact allow us to contrast results obtained for the SPP model to those of other well-studied percolation models. For $C=1$, the SPP model effectively reduces to the ordinary bond-percolation model on ER graphs displaying a smooth transition when a fraction $p_c = 1 - 1/\bar{k}$ of randomly selected edges is removed from the graph. For $1 < C \leq N$, the SPP model differentiates from the ordinary bond-percolation model as edges in the graph are no longer deleted independently, rather in a correlated fashion (note that $C=N$ is a limiting case, as the inequality $Q_t \leq C$ always holds as long as $o_t$ and $d_t$ are in the same connected component of the graph $\mathcal{G}_t$). We explicitly refer to the infinite-$C$ SPP model when $\lim_{N \to \infty} C = \infty$; the finite-$C$ SPP model occurs otherwise.



\begin{table}[!htb]
\renewcommand{\arraystretch}{1.2}
\centering
 \begin{tabular}{cccc}
 \hline
 \hline
 $C$ & $p_{c}$ & $\beta/\bar{\nu}$ & $\gamma/\bar{\nu}$  \\ 
 \hline
 $1$ & $0.750 \pm 0.001$ & $0.32\pm0.01$ & $0.34 \pm 0.01$ \\ 
 $2$ & $0.701 \pm 0.001$ & $0.32 \pm 0.01$ & $0.34 \pm 0.01$ \\
 $3$ & $0.682 \pm 0.001$ & $0.32 \pm 0.01$ & $0.34 \pm 0.01$  \\
 $\infty$ & $0.646\pm 0.001$ & $0.21 \pm 0.01$ & $0.55 \pm 0.01$ \\
 \hline
 \hline 
\end{tabular}
\caption{Critical properties of the shortest-path-percolation model. From left to right, we report: the value of the input parameter $C$ of the SPP model, the estimate of the critical threshold $p_c$, the ratio of the critical exponents $\beta/\bar{\nu}$ and $\gamma/\bar{\nu}$. Estimates reported here are obtained via finite-size scaling analysis in the event-based ensemble.} 
\label{tab:summary}
\end{table}

We fully characterize the behavior of the SPP model on ER graphs with a systematic numerical analysis.  
Our results are based on a large number of independent simulations 
for each combination of $N$ and $C$ values, see Supplemental Material (SM) for details. In each realization, we first generate an ER graph with average degree $\bar{k} = 4$ and then apply the SPP model to it. For finite $C$, the computational complexity of the SPP model scales slightly super-linearly with the system size, so we  consider networks with size up to $N=2^{27}$. Simulating the infinite-$C$ SPP model is subject to a higher computational burden, hence we analyze networks with size up to $N=2^{20}$.
As for the control parameters, we focus our attention to both the fraction of removed edges $p$ as well as the raw number of demanded origin-destination pairs $t$. 
We determine the properties of the SPP transition via finite-size scaling (FSS) analysis relying on the conventional ensemble where sampled configurations correspond to independent realizations of the SPP model obtained at specific values of the control parameters~\cite{stauffer2018introduction}. All results hold when using the so-called event-based ensemble~\cite{fan2020universal, li2023explosive}. To construct this ensemble, we still sample one configuration  from each individual realization of the SPP model; such a sampled configuration is the one corresponding to the largest change, caused by the deletion of a single edge, in the size of the largest cluster during the SPP process.

Our main finding is that SPP belongs to the same universality class as of ordinary percolation as long as $C$ is finite; for infinite $C$, the SPP transition becomes more abrupt than ordinary percolation, being characterized by a set of different critical exponents, see Table~\ref{tab:summary} and SM. 
Results for the finite-$C$ class are obtained by setting $C=1, 2$ and $3$ (main text and SM); for the infinite-$C$ class, our findings are obtained for $C=N$ (main text and SM) as well as for $C=N^{1/3}$ and $C = \log{(N)}$ (SM).

\begin{figure}[!htb]
    \centering
    \includegraphics[width=\columnwidth]{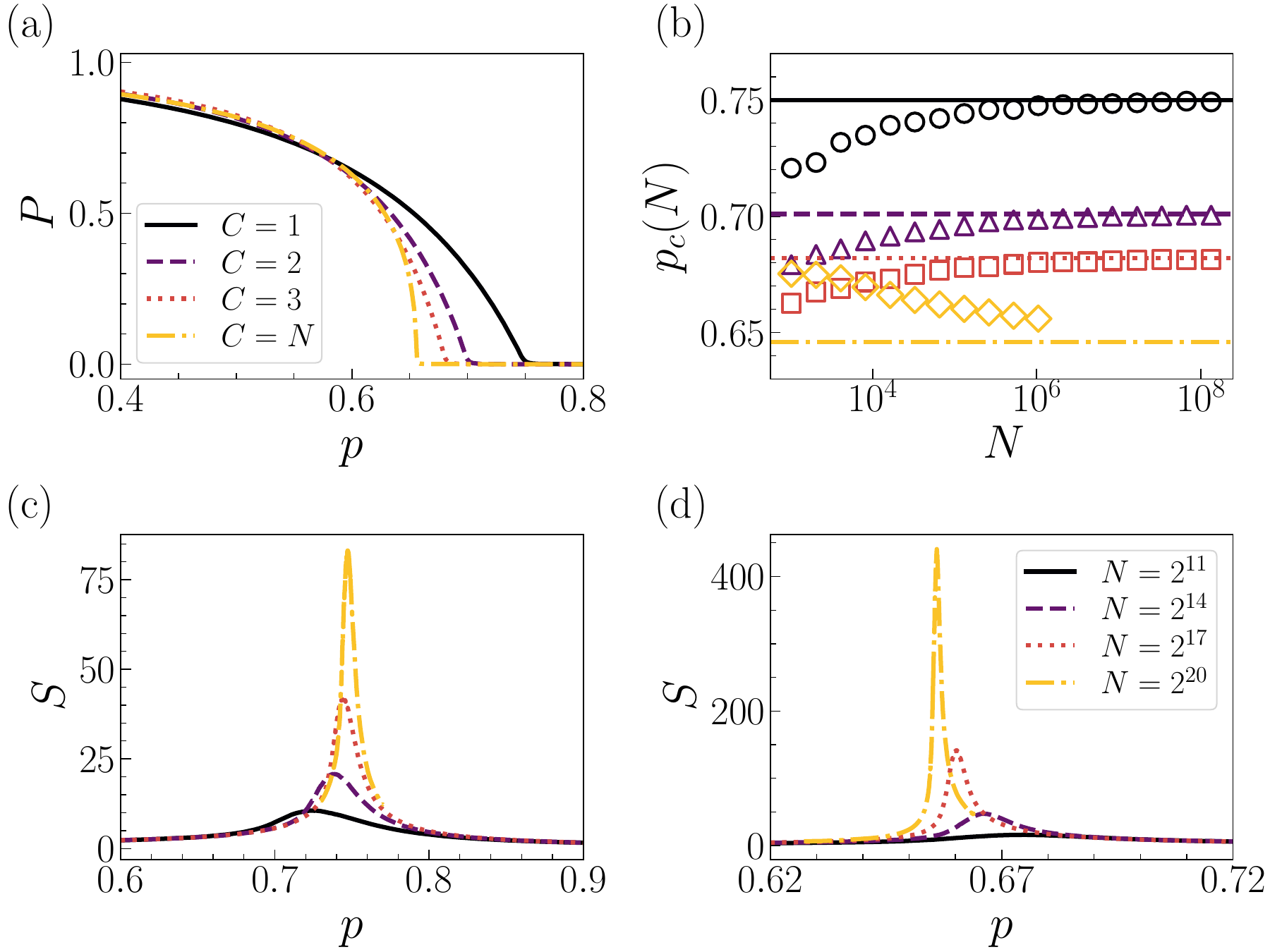}
    \caption{
    Shortest-path percolation transition in the conventional ensemble. 
    (a) Percolation strength $P$ as a function of fraction of removed edges $p$ with different values of $C$. The SPP model is applied to Erd\H{o}s-R\'enyi (ER) graphs with size $N=2^{20}$.
    (b) Pseudocritical point $p_c(N)$ as a function of $N$ for different values of $C$. The horizontal line denotes the estimated critical point $p_c$ for each $C$. (c) Average cluster size $S$ as a function of $p$ for ER graphs 
    for different network sizes $N$. Here $C=1$. (d) Similar to (c) but for $C=N$.  
    \label{figure2}
    }
\end{figure}

\begin{figure}[!htb]
    \centering
    \includegraphics[width=\columnwidth]{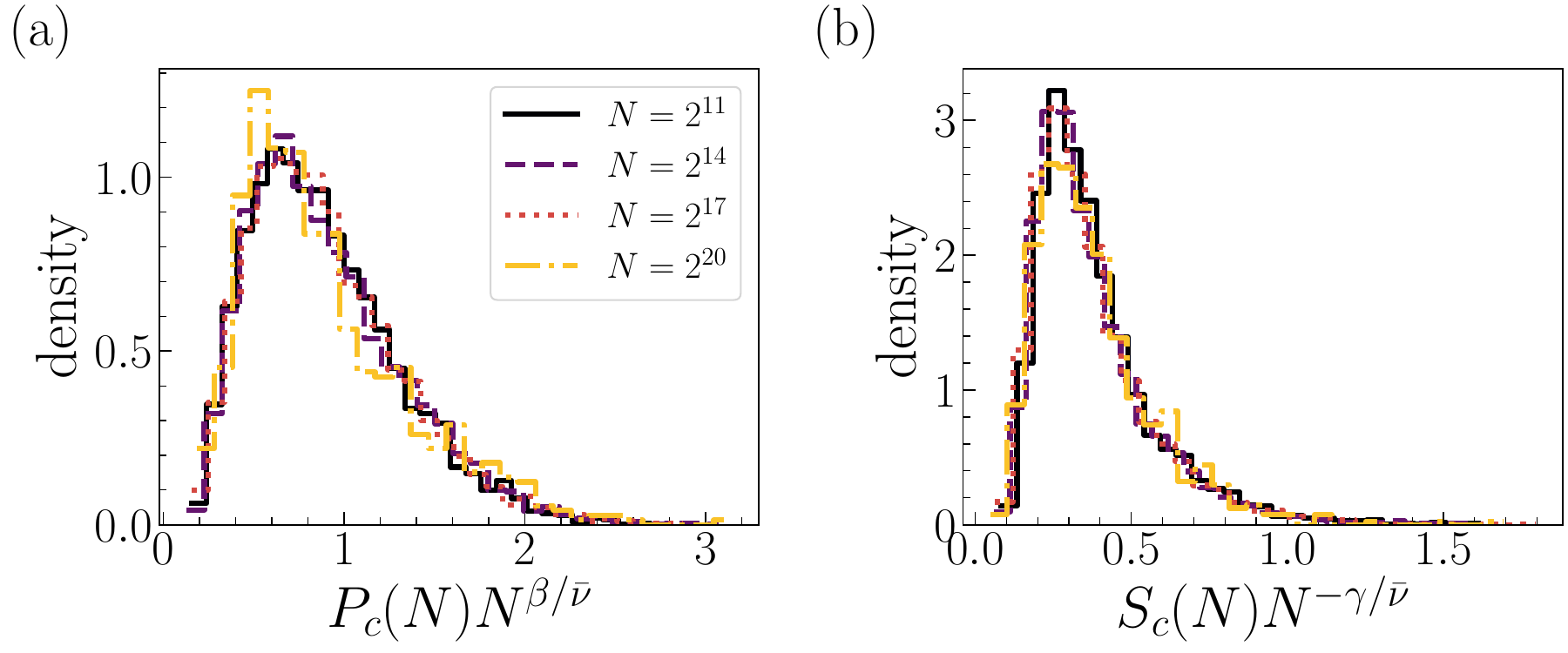}
    \caption{Shortest-path-percolation transition in the event-based ensemble. Here we use $C=N$.
    (a) The distribution of ${P}_{c}(N) \, N^{\beta/\bar{\nu}}$ with $\beta/\bar\nu = 0.21$.
    Different curves correspond to different network sizes $N$. (b) Same as in (a) but for  $S_{c}(N) \,N^{-\gamma/\bar{\nu}}$ with $\gamma/\bar\nu = 0.55$.
    }
    \label{figure3}
\end{figure}

In \FIG{figure2}, we report results valid under the conventional ensemble. In \FIG{figure2} (a), we plot the percolation strength 
$P$ as a function of $p$. We fix the network size to $N=2^{20}$ and compare results obtained for different $C$ values.
We observe that, as $C$ increases, the change displayed by $P$ becomes more abrupt. The different behaviors of the 
finite- vs infinite-$C$ cases
are apparent by looking at how the average cluster size $S$ changes as a function of $p$, see \FIGS{figure2} (c) and \ref{figure2} (d), with $S$ being characterized by a peak value $S_c(N)$ occurring at the pseudocritical point $p_c(N)$. 
We observe that $S_c(N) \sim N^{\gamma/\bar{\nu}}$, with $\gamma/\bar{\nu} = 0.35 \pm 0.01$ if $C=1$ and $\gamma/\bar{\nu} = 0.53 \pm 0.01$ for $C=\infty$ (see SM). 
Also, we find that $p_c = p_c(N)  + b N^{-1/\bar{\nu}}$ for any $C$ value, see \FIG{figure2} (b). Not surprisingly, the value of the critical point $p_c$ is a decreasing function of $C$, ranging from $p_c=0.750 \pm 0.001$ for $C=1$ to $p_c = 0.646 \pm 0.001$ for $C=\infty$; 
however, we surprisingly find that $b>0$ and $1/\bar{\nu} \simeq 1/3$ for finite $C$, but $b < 0$ and $1/\bar{\nu} = 0.18 \pm 0.01$ for infinite $C$.  
The observed difference  
in the value of the critical exponent $\bar{\nu}$ 
as well as
the change of the sign of the fitting parameter $b$ denote that a fundamentally different type of percolation transition takes place depending on whether $C$ is finite or infinite.

In \FIG{figure3}, we display results for the FSS analysis under the event-based ensemble. Specifically, we display the collapse of the distributions of the rescaled pseudocritical observables $P_c(N) \, N^{\beta/\bar{\nu}}$ and $S_c(N) \, N^{-\gamma/\bar{\nu}}$.
Both plots appearing in \FIG{figure3} refer to the infinite-$C$ case; we report results valid for finite $C$ in the SM.
We find $\beta/\bar{\nu} = 0.21 \pm 0.01$ and $\gamma/\bar{\nu} = 0.55 \pm 0.01$ for $C=\infty$.
Comparable values of the ratio $\beta/\bar{\nu}$ are obtained by monitoring the scaling of the peak values of the $k$th largest clusters, for $k=2, 3, 4,$ and $5$ (see SM).
Note that these estimates are compatible with the known hyperscaling relation $2 \beta/\bar{\nu} + \gamma/\bar{\nu} = 1$. 
For finite $C$, we  recover $\beta/\bar{\nu} \simeq \gamma/\bar{\nu} \simeq 1/\bar{\nu} \simeq 1/3$ for both the ensembles, as expected for ordinary percolation~\cite{stauffer2018introduction} (see SM).
The critical exponent ratios $\beta/\bar{\nu}$ and $\gamma/\bar{\nu}$ obtained via FSS at the pseudocritical point $p_c(N)$ in the conventional ensemble are compatible with those valid for the event-based ensemble; at the critical point $p_c$, the estimates are different, likely because affected by finite-size effects (SM).

\begin{figure}[!htb]
    \centering
    \includegraphics[width=\columnwidth]{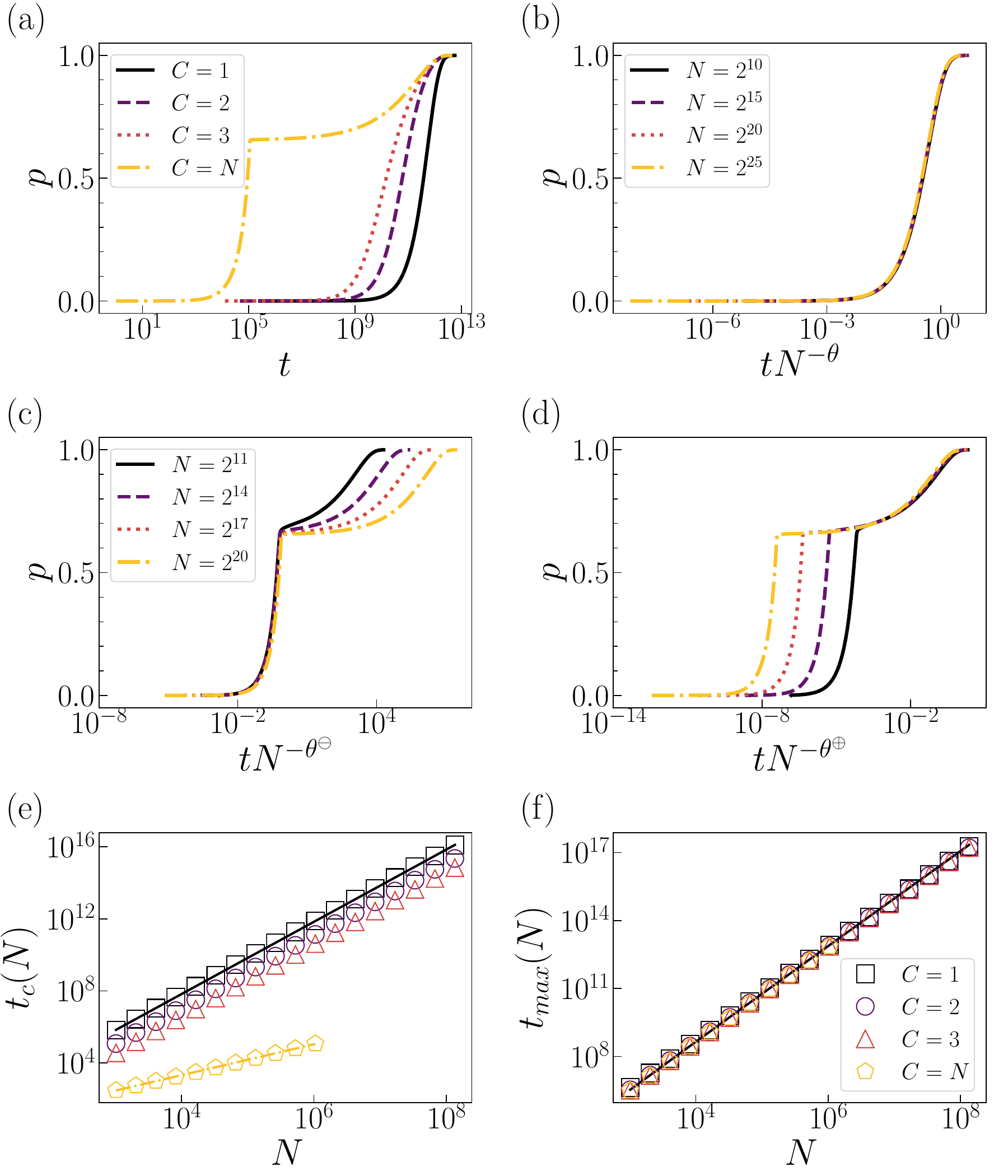}
    \caption{
    Control parameters in the shortest-path percolation model.
    (a) The fraction of removed edges $p$ is plotted as a function of number of demanding agents $t$ for different values of $C$. Results are valid for Erd\H{o}s-R\'enyi graphs with size $N=2^{20}$.
    (b) Curve collapse for $C=1$ and different $N$ values. The abscissas values are rescaled as $t N^{-\theta}$, with $\theta = 2.01$, to obtain a collapse between the various curves. (c) Same as in (b), but for $C=N$. The  collapse is obtained by rescaling the abscissas as $t N^{-\theta^{\ominus}}$, with $\theta^{\ominus} = 0.86$. (d) Same as in (c), but rescaling the abscissas as $t N^{-\theta^{\oplus}}$, with $\theta^{\oplus} = 2.07$.
    (e) Pseudocritical threshold $t_c(N)$ as a function of $N$ for different values of $C$. The full black line indicates the scaling  $t_{c}(N) \sim N^{2}$, whereas the dashed yellow line stands for  $t_{c}(N) \sim N^{0.86}$.
    (f) Number of demanded pairs required to fully dismantle the network $t_{max}(N)$ {\it vs.} $N$ for different values of $C$. The black line indicates the scaling  $t_{\max}(N) \sim N^{2}$.
}

    \label{figure4}
\end{figure}

The study of the standard observables of \FIG{figure2} and \FIG{figure3} highlights a marked difference
between the finite- and infinite-$C$ transitions, however, does not convey an actual physical explanation of such a finding. A clear picture emerges from the analysis of
\FIG{figure4}, where we study the mapping between the two natural control parameters of the SPP model, i.e., $p$ and $t$.
For finite $C$, $p$ and $t$ are mapped one to the other by a universal smooth function that is revealed by rescaling $t \mapsto t N^{-\theta}$. We estimate $\theta = 2.01 \pm 0.01$ for $C=1$ and similar values for other finite-$C$ cases, see \FIG{figure4}(e) and SM. The scaling exponent $\theta \simeq 2$ tells us that the dismantling of the network requires to select a number of origin-destination pairs $t_{\max}(N)$ that is proportional to the total number of node pairs in the network, see \FIG{figure4}(f). When $C$ is infinite, however, two distinct scaling behaviors are visible: (i) for $t \leq t_c(N)$, curve collapse is obtained by plotting $p$ vs $t N^{-\theta^{\ominus}}$ with $\theta^{\ominus} = 0.86 \pm 0.01$, \FIG{figure4}(e) and SM; (ii) for $t > t_c(N)$, data collapse is obtained by rescaling $t \mapsto t N^{-\theta^{\oplus}}$ with $\theta^{\oplus} = 2.07 \pm 0.01$, \FIG{figure4}(f) and SM. Note that the kink point $t_c(N)$ is such that $p(t_c(N)) = p_c(N)$. 
The finding can be interpreted as follows. When the giant cluster is present, edges are removed with ease as a path exists between most pairs of nodes. At the beginning, the number of removed edges per pair is small as the shortest-path length between pairs of randomly selected nodes is short. However, as the graph becomes sparser, the average shortest-path length increases, so does the number of edges removed per demanded pair. Critical behavior corresponds to the consumption of a large number of edges for a small number of demanded pairs. In particular, the number of pairs that needs to be demanded to reach the critical point is a vanishing fraction of the total number of pairs of nodes in the network, as the value of the scaling exponent $\theta^{\ominus}$ indicates. After such a massive consumption of edges, the network is fragmented into multiple clusters. In this configuration, two randomly selected nodes are unlikely to belong to the same connected component. This leads to a dramatic slow down in the number of edges removed per pair of demanded origin-destination nodes. Also, individual clusters have a finite diameter, thus each of them is dismantled by an effectively finite-$C$ SPP process, hence $\theta^{\oplus} \simeq 2$.

\begin{figure}[!htb]
    \centering
    \includegraphics[width=\columnwidth]{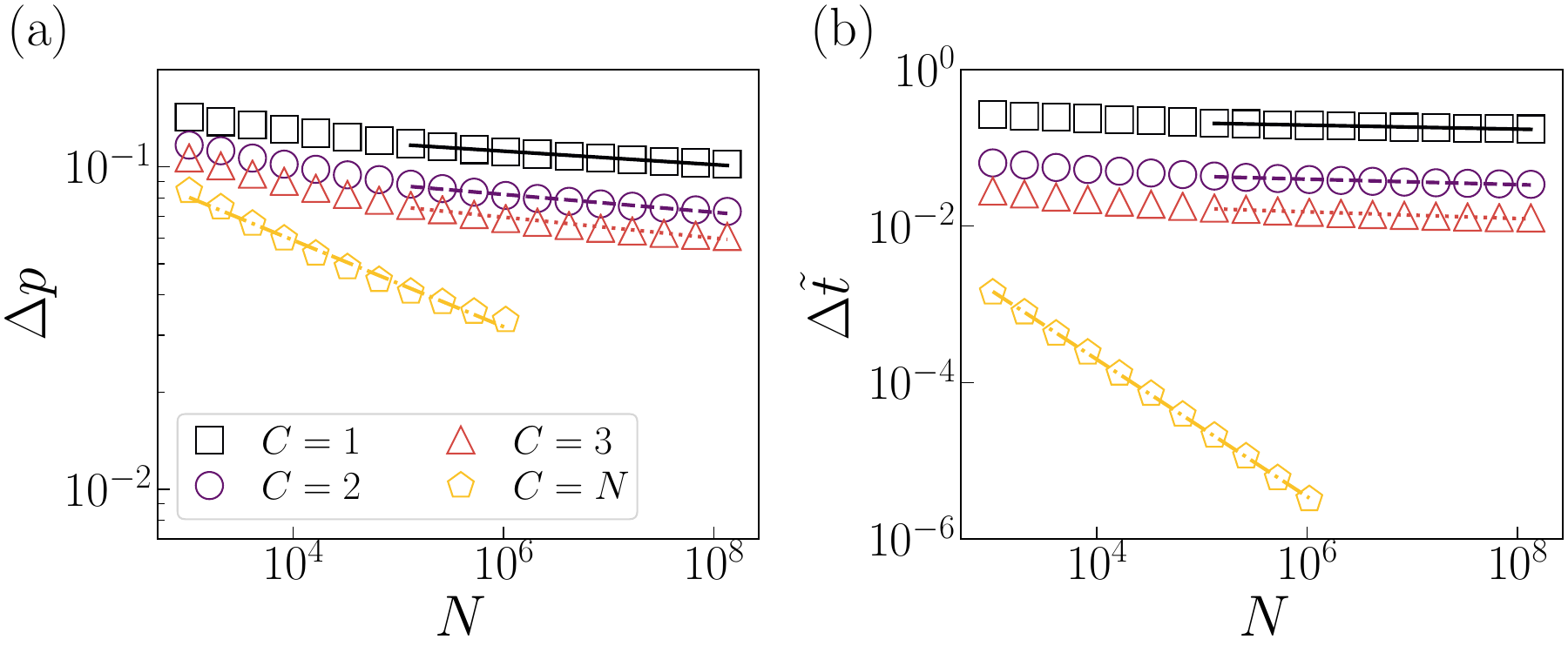}
    \caption{
    Abruptness of the shortest-path-percolation transition.
    (a) We plot $\Delta p$ {\it vs.} $N$ for different $C$ values. Displayed lines represent the best fits of the scaling $\Delta p \sim N^{-\alpha}$. We get $\alpha = 0.02 \pm 0.01$ for $C=1$, $\alpha = 0.03 \pm 0.01$ for $C=2$, $\alpha = 0.03 \pm 0.01$ for $C=3$, and $\alpha = 0.13 \pm 0.01$ for $C=N$. (b) $\Delta \tilde{t}$ {\it vs.} $N$. Lines stand for best fits of the scaling $\Delta \tilde{t} \sim N^{- \alpha'}$. We find $\alpha' = 0.03 \pm 0.01$ for $C=1$, $\alpha' = 0.03 \pm 0.01$ for $C=2$, $\alpha' = 0.04 \pm 0.01$ for $C=3$, and $\alpha' = 0.88 \pm 0.01$ for $C=N$.
    }
    \label{figure5}
\end{figure}

We assess the abruptness of the SPP transition using the same procedure as in Refs.~\cite{achlioptas2009explosive, nagler2011impact}, and measure the width of the transition window as $\Delta p = p_1 - p_2$ or $\Delta \tilde{t} = (t_1 - t_2) \, N^{-2}$, where $p_2$ and $t_2$ are the highest values of the control parameters for which $P > 0.5$, whereas $p_1$ and $t_1$ are the highest values of the control parameters for which $P > 1 / \sqrt{N}$. The width of the transition in terms of origin-destination pairs of nodes is further normalized. As the results of \FIG{figure5} show, we find that $\Delta p \sim N^{-\alpha}$. The value of the scaling exponent is almost zero for finite $C$, e.g.,  $\alpha = 0.02 \pm 0.01$ for $C=1$, denoting that the transition is continuous; for $C = N$,
we find $\alpha = 0.13 \pm 0.01$, meaning that the transition is weakly discontinuous~\cite{nagler2011impact}. 
In the latter case, the exponent value is smaller than the one observed for the explosive percolation transition, thus indicating a less abrupt change of phases. The weakly discontinuous nature of the infinite-$C$ SPP transition becomes very apparent by looking at the scaling $\Delta \tilde{t} \sim N^{-\alpha'}$, where we find $\alpha' = 0.88 \pm 0.01$. 
We find instead $\alpha' \simeq 0$ for finite $C$, e.g., $\alpha' = 0.03 \pm 0.01$ for $C=1$, once more denoting a continuous transition.


To sum up, we introduce the shortest-path-percolation model aimed at mimicking the utilization, and eventual exhaustion, of a network's resources by agents demanding minimum-cost itineraries below a certain budget. The main finding of our systematic analysis about the application of the SPP model to Erd\H{o}s-R\'{e}nyi graphs is that, if budget is finite, then exhaustion occurs like in an ordinary percolation process; however, if budget is unbounded, then the network's resources are consumed abruptly, in a similar fashion as for the of explosive percolation model~\cite{achlioptas2009explosive}. 
However, the abruptness of the SPP transition is not the result of a competitive selection criterion that takes advantage of knowledge about the cluster structure of the graph as in the explosive percolation model~\cite{nagler2011impact}, rather caused by topological correlations among groups of deleted edges.
Our findings underscore that not only dynamical processes, but also fundamental structural transitions such as percolation are radically altered by framing them in terms of path-based rather than edge-based models~\cite{lambiotte2019networks}. Also, they provide further evidence about the plausibility of range-dependent universality classes in network percolation~\cite{almeira2020scaling, almeira2021explosive,cirigliano2023extended}.

The SPP model can be easily adapted to deal with arbitrary forms of demand and cost functions. Also, it can be naturally extended to directed, weighted, time-stamped graphs. All these generalizations are necessary to make the model useful for the development of computational frameworks aimed at analyzing and optimizing real-world infrastructural networks.


\begin{acknowledgements}
    The authors thank G. Bianconi and S. Dorogovtsev for comments on the initial stages
    of this project. The authors acknowledge support by the Army Research Office under contract number W911NF-21-1-0194 and by the Air Force Office of Scientific Research under award number FA9550-21-1-0446. The funders had no role in study design, data collection, and analysis, the decision to publish, or any opinions, findings, conclusions, or recommendations expressed in the manuscript.
\end{acknowledgements}


\bibliography{apssamp}


\clearpage
\onecolumngrid

\renewcommand{\theequation}{S\arabic{equation}}
\setcounter{equation}{0}
\renewcommand{\thefigure}{S\arabic{figure}}
\setcounter{figure}{0}
\renewcommand{\thetable}{S\arabic{table}}
\setcounter{table}{0}

\section{Supplemental Material}

\section{The shortest-path-percolation model on Erd\H{o}s-R\'{e}nyi graphs}

All the results of the present paper are based on the application of the shortest-path-percolation (SPP) model to Erd\H{o}s-R\'{e}nyi (ER) graphs. Graphs $\mathcal{G}_1$ used as inputs of the SPP algorithm are  generated by drawing exactly $E_1 = N \bar{k} / 2$ edges between pairs of randomly chosen nodes. Here $N$ is the size of the graph, and $\bar{k}$ indicates the average degree of the graph. We use $\bar{k} = 4$ in all our simulations, although the choice affects only the location of the critical points and not the other critical properties of the SPP transition. In our simulations, pairs of origin-destination nodes are also extracted at random from all possible nodes in the graph. We extract random pairs of origin-destination nodes until all edges in the graph are removed.

The output of each realization of the SPP model is an ordered list containing all edges of the input graph $\mathcal{G}_1$. These edges appear in the order in which they are removed according to the specific realization of the SPP model. Each edge has also associated its corresponding $t$ value denoting the index of the  origin-destination pair requested during the application of the SPP model. We use such an output list in all our subsequent analyses. To efficiently keep track of the cluster structure of the network during the SPP process, we take advantage of the Newman-Ziff algorithm~\cite{newman2000efficient}.

Results are obtained on a large number of numerical simulations for each combination of the parameter $C$ of the SPP model and size of the network $N$, see Table~\ref{tab:SPP_summary} for details.

\begin{table}[!htb]
\renewcommand{\arraystretch}{1.2}
\centering
 \begin{tabular}{rrrrr}
 \hline
 \hline
 $N$ & $C=1$ & $C=2$ & $C=3$ & $C=N$  
 \\ 
 \hline
$2^{10} - 2^{17}$ & $5,000$ & $5,000$ & $5,000$ & $5,000$
\\
$2^{18} - 2^{20}$ & $1,000$ & $1,000$ & $1,000$ & $1,000$
\\
$2^{21} - 2^{25}$ & $1,000$ & $1,000$ & $1,000$ & -
\\
$2^{26}$ & $200$ & $200$ & $200$ & -
\\
$2^{27}$ & $20$ & $20$ & $20$ & -
\\
 \hline
 \hline 
\end{tabular}
\caption{Numerical simulation of the shortest-path-percolation (SPP) model. For each size $N$ of the network and value of the cost $C$, we report the corresponding number of numerical simulations that we performed.} 
\label{tab:SPP_summary}
\end{table}

\section{Control and order parameters}


The natural control parameter of the shortest-path-percolation (SPP) model is $t$, i.e., the number of demanded pairs of origin-destination nodes. Another control parameter is the fraction of removed edges, namely $p$. This control parameter allows for immediate comparisons between the SPP model and other standard percolation models. The fraction of removed edges is associated to $t$ as
\begin{equation}
p (t) = 1 - \frac{E_t}{E_1} \; ,
\label{eq:p_t_map}
\end{equation}
where $E_t = | \mathcal{E}_t |$ is the size of the set of edges associated to the graph $\mathcal{G}_t$. 
Specifically, $\mathcal{G}_{t} = (\mathcal{V}, \mathcal{E}_{t})$ is the graph available when the $t$-th pair of origin-destination nodes is demanded, composed of $N = |\mathcal{V}|$ nodes and $E_t= |\mathcal{E}_t|$ edges.


We consider two standard metrics in network percolation: the percolation strength $P$ and the average cluster size $S$. These metrics serve to characterize the cluster structure of a network under the SPP model.

Indicate with $s_1 \leq s_2 \leq \cdots \leq s_{R-1} \leq s_R$ the size of the $R$ connected components or clusters observed in the network at a certain stage of the SPP model. We clearly have that $\sum_{r=1}^R s_r = N$, i.e., the sum of all clusters sizes equals the network size $N$.

$P$ is defined as the size of the largest cluster divided by the network size, thus
\begin{equation}
P = \frac{s_R}{\sum_{r=1}^R s_r} \; ,
    \label{eq:perc_str}
\end{equation}
$S$ is instead defined as
\begin{equation}
S = \frac{\sum_{r=1}^{R-1} s_r^2}{\sum_{r=1}^{R-1} s_r} \; ,
    \label{eq:av_cluster}
\end{equation}
where both the above sums runs over all clusters except the largest one.

\section{Power-law fit}

In our analysis, we systematically rely on power-law fits of data.
We start from the hypothesis that variables $x$ and $y$ are related by
\begin{equation}
x = z \,  y^{\zeta} \; .
\label{eq:power}
\end{equation}
To determine the best values of $z$ and $\zeta$, we first take the logarithms of both sides of Eq.~(\ref{eq:power}) to obtain

\begin{equation}
\log (x) = \log(z) + \zeta \, \log(y)\; ,
\label{eq:log_power}
\end{equation}
and then apply the simple linear regression algorithm. Such an algorithm provides us with the best estimates $\hat{(\log z)}$ and $\hat{\zeta}$, their associated errors $d (\log z)$ and $d \zeta$ (i.e., standard errors under the assumption of residual normality), as well as the metric of goodness of the fit $r$ (i.e., Pearson correlation coefficient). Our estimates of the exponent are in the format $\zeta = \hat{\zeta} \pm d \zeta$. 
For consistency with previous work, we report numerical values of these quantities with a precision of two decimals~\cite{fan2020universal}. We compare $r$ values to determine if one power-law scaling is better than another in some of our analyses. 

\begin{figure}[!htb]
    \centering
    \includegraphics[width=0.8\columnwidth]{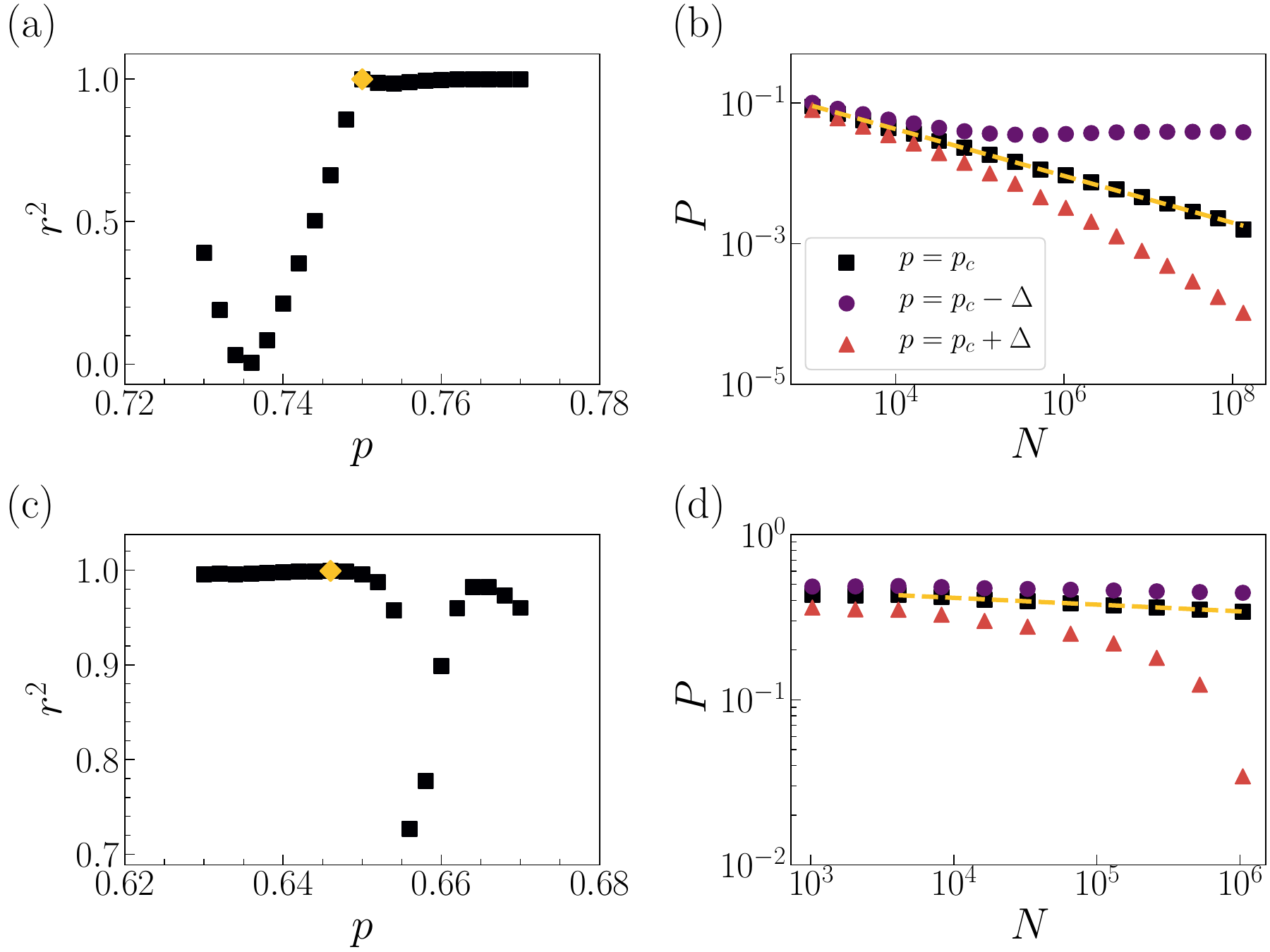}
    \caption{
    Estimation of the critical threshold in the conventional ensemble.
    (a) Coefficient of determination $r^2$ as a function of $p$. Here $C=1$. (b) Plot of $P$ as a function of $N$ at $p=p_c$, $p=p_c-\Delta$, and $p=p_c+\Delta$, respectively. We use $p_c=0.75$ and $\Delta=0.005$. Here also $C=1$. (c) Similar to (a) but for $C=\infty$. (d) Similar to (b) but for $C=\infty$. We use $p_c=0.646$ and $\Delta=0.01$. Detailed results can be found in Table~\ref{tab:FSS_conventional_full}.}
    \label{fig:SM-power-law-fit}
\end{figure}

For instance, in \FIG{fig:SM-power-law-fit} (a), we plot the $r^2$, i.e., the coefficient of determination of simple linear regression, as a function of $p$ while estimating $p_c$ of the SPP model with $C=1$ by using the power-law fit. The yellow diamond symbol denotes the point where $r^2$ is maximized, $p_c=0.75$. In \FIG{fig:SM-power-law-fit} (b), we plot $P$ as a function of $N$ at $p=p_c$, $p=p_c-\Delta$, and $p=p_c+\Delta$, respectively. Here, we use $p_c=0.75$ and $\Delta=0.005$. We can clearly observe that $p=p_c$ provides the best power-law fit.

\section{Finite-size scaling analysis}

\subsection{Conventional ensemble}

We denote with $P(p, N)$ and $S(p, N)$ the values of the metrics of Eqs.~(\ref{eq:perc_str}) and~(\ref{eq:av_cluster}), respectively, when exactly a fraction $p$ of edges is removed from the graph.
To quantify the critical properties of the SPP transition, we rely on finite-size scaling (FSS) analysis~\cite{stauffer2018introduction} based on the scaling ansatzs 
\begin{equation}
    \langle P (p, N) \rangle = N^{-\beta/\bar\nu} F^{(1)}[(p-p_c)N^{1/\bar\nu}]
    \label{eq:FSS_ansatz1}
\end{equation}
and
\begin{equation}
    \langle S (p, N) \rangle = N^{\gamma/\bar\nu} F^{(2)}[(p-p_c)N^{1/\bar\nu}] \; ,
    \label{eq:FSS_ansatz2}
\end{equation}
where $\langle P(p, N) \rangle$ and $\langle S(p, N) \rangle$ are the ensemble averages of $P(p, N)$ and $S(p, N)$, respectively, 
$F^{(1)}(\cdot)$ and $F^{(2)}(\cdot)$ are universal scaling functions, while $\beta$, $\gamma$, and $\bar\nu$ are critical exponents for the percolation strength, the average cluster size, and the correlation volume, respectively. 
In Eqs.~(\ref{eq:FSS_ansatz1}) and~(\ref{eq:FSS_ansatz2}), $p_c$ is the critical value of the control parameter where the transition occurs in a network of infinite size. In a network of finite size $N$, the pseudo-critical threshold $p_c(N)$ indicates the value of the control parameter related to the biggest variation in the cluster structure of the network. There are multiple ways of defining this condition. We use
\begin{equation}
    p_{c}(N) = \arg \max_p \, \left\langle S(p, N) \right\rangle \; .
    \label{eq:conv_pseudo_threshold}
\end{equation}
Still according to FSS, critical and pseudo-critical points are related by 
\begin{equation}
    p_{c} = p_{c}(N) + b N^{-1/\bar\nu} ,
    \label{eq:pseudo_scaling}
\end{equation}
where $b$ is a constant.

\subsection{Critical properties of the shortest-path-percolation in the conventional ensemble}

Note that we use a simplified notation to denote  $\langle P(p, N) \rangle$ and $\langle S(p, N) \rangle$ in the figures, and in the text. Details on the notation are provided below.

\begin{figure}[!htb]
    \centering
    \includegraphics[width=\columnwidth]{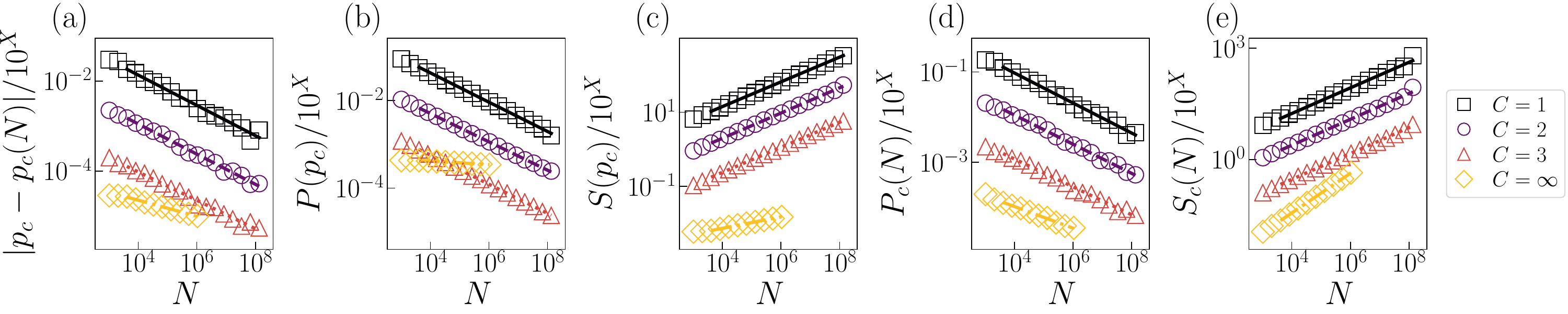}
    \caption{
    Finite-size scaling analysis in the conventional ensemble.
    An additional factor $10^X$ is used for visual clarity. We use $X=0, 1, 2, 3$ for $C=1, 2, 3, N$, respectively. In each panel, we display as symbols the average value of the metric. Error bars are not displayed 
    to avoid overcrowding the figure.
    Lines appearing in the plot are best fits obtained via simple linear regression 
    on log-transformed variables, see text for details.
    (a) Scaling of Eq.~(\ref{eq:pseudo_scaling}) to estimate $1/\bar\nu$. (b) Scaling at the critical point for the percolation strength, i.e, $P(p=p_c)$ {\it vs} $N$, to estimate $\beta/\bar\nu$. (c) Scaling at the critical point for the average cluster size, i.e, $S(p=p_c)$ {\it vs} $N$, to estimate $\gamma/\bar\nu$. (d) Scaling at the pseudo-critical point for the percolation strength, i.e., $P(p=p_c(N))$ {\it vs} $N$, to estimate $\beta/\bar\nu$. (e) Scaling at the pseudo-critical point for the average cluster size, i.e., $S(p=p_c(N))$ {\it vs} $N$, to estimate $\gamma/\bar\nu$. Estimated exponents are reported in Table~\ref{tab:FSS_conventional_full}.}
    \label{fig:SM-FSS-convention}
\end{figure}

We determine the value of the critical threshold $p_c$ using Eq.~(\ref{eq:FSS_ansatz1}). For $p = p_c$, the FSS ansatz predicts 
\begin{equation}
P(p_c) = \langle P (p=p_c, N) \rangle \sim N^{-\beta/\bar\nu} \; ,
\label{eq:perc_str_critical}
\end{equation}
with $P(p_c)$ highlighting the simplified notation adopted in the corresponding figure. Thus, we obtain the best estimator of the critical threshold as the value of the control parameter $p$ leading to the best power-law fit, \FIG{fig:SM-FSS-convention}. The best estimate of the $\beta/\bar\nu$ is obtained from the very same power-law fit. The ansatz of Eq.~(\ref{eq:FSS_ansatz2}) together with the estimated $p_c$ value further allows us to obtain the best estimate of $\gamma/\bar\nu$, i.e., $S(p_c) = \langle S (p=p_c, N) \rangle \sim N^{\gamma/\bar\nu}$, see \FIG{fig:SM-FSS-convention}(c). 

For finite $C$, these scalings provide good estimates of the critical point and of the critical exponents. For infinite $C$, the estimates of the critical exponents are instead affected by strong finite size effects  while the estimate of the critical point remains unaffected.

Also, we estimate $\beta/\bar\nu$ and $\gamma/\bar\nu$  using the scaling 
\begin{equation}
P_c(N) = \langle P(p=p_c(N), N) \rangle \sim N^{-\beta/\bar\nu} \;,
\label{eq:perc_str_ps_critical}
\end{equation}
\begin{equation}
S_c(N) =  \langle S(p=p_c(N), N) \rangle \sim N^{\gamma/\bar\nu} \; , 
\label{eq:average_ps_critical}
\end{equation}
as predicted by Eqs.~(\ref{eq:FSS_ansatz1}),~(\ref{eq:FSS_ansatz2}) and~(\ref{eq:pseudo_scaling}). $p_c(N)$ is defined in Eq.~(\ref{eq:conv_pseudo_threshold}). See \FIG{fig:SM-FSS-convention}(d) and (e) for details. These provide us with accurate estimates of the critical exponents for any $C$ value.

Finally, we perform the power-law fit of the expression $p_c - p_c(N) = b N^{-1/\bar\nu}$, see Eq.~(\ref{eq:pseudo_scaling}), 
to determine the exponent $1/\bar\nu$ as well as the value of the constant $b$, \FIG{fig:SM-FSS-convention}(a). Numerical estimates of the critical parameters are reported in Table~\ref{tab:FSS_conventional_full}.

\begin{table}[!htb]
\renewcommand{\arraystretch}{1.2}
\centering
\begin{tabular}{ccc|cc|cc}
 \hline
 \hline
 \multicolumn{3}{c|}{} &  \multicolumn{2}{c|}{critical} & \multicolumn{2}{c}{pseudo-critical}\\
 \hline
 \hline
 $C$ & $p_{c}$ & $1/\bar\nu$ & 
 $\beta/\bar{\nu}$ & $\gamma/\bar{\nu}$  & $\beta/\bar{\nu}$ & $\gamma/\bar{\nu}$\\ 
 \hline
 1 & $0.750 \pm 0.001$ & $0.34 \pm 0.01$ &$0.33 \pm 0.01$ & $0.33 \pm 0.01$ & $0.33 \pm 0.01$ & $0.35 \pm 0.01$\\ 
 2 & $0.701 \pm 0.001$ & $0.34 \pm 0.01$ & $0.32 \pm 0.01$ & $0.34 \pm 0.01$ & $0.32 \pm 0.01$ & $0.35 \pm 0.01$\\
 3 & $0.682 \pm 0.001$ & $0.32 \pm 0.01$ & $0.33 \pm 0.01$ & $0.34 \pm 0.01$ & $0.31 \pm 0.01$ & $0.35 \pm 0.01$\\
 $\infty$ & $0.646 \pm 0.001$ & $0.18 \pm 0.01$ & $0.04 \pm 0.01$ & $0.15 \pm 0.01$ & $0.24 \pm 0.01$ & $0.53 \pm 0.01$\\
 \hline
 \hline 
\end{tabular}
\caption{Critical properties of the shortest-path-percolation (SPP) model. Estimates reported here are obtained via finite-size scaling analysis in the conventional ensemble.} 
\label{tab:FSS_conventional_full}
\end{table}

\subsection{Event-based ensemble}

Also here, we denote with $P(p, N)$ and $S(p, N)$ the values of the metrics of Eqs.~(\ref{eq:perc_str}) and~(\ref{eq:av_cluster}), respectively, when exactly a fraction of $p$ edges is removed from the graph. The single-instance pseudo-critical point can be defined as in Ref.~\cite{fan2020universal}, i.e.,
\begin{equation}
    p^{+}(N) =  \arg \max_p \, \left[ P(p - 2/N \bar{k}, N) - P(p, N)  \right]  \; ,
    \label{eq:ev_pseudo_threshold1}
\end{equation}
or as in Ref.~\cite{li2023explosive}, i.e.,
\begin{equation}
    p^{-}(N) =  \arg \max_p \, \left[ P(p, N) - P(p + 2/N \bar{k} , N) \right] \;  .
    \label{eq:ev_pseudo_threshold2}
\end{equation}
In the above equations, the superscript $+$ ($-$) indicates whether the value of the control parameter $p$ is measured immediately after (before) the largest drop, induced by the removal of a single edge, in the value of the order parameter $P$
 is observed. Note that the values $p^{+}(N)$ and $p^{-}(N)$ are measured for each individual realization of the SPP model.
For such an individual realization, we can also measure the
single-instance
pseudo-critical value of the percolation strength, i.e.,
 \begin{equation}
    P^{+}(N) =  P (p = p^{+}(N), N)  \; .
    \label{eq:ev_pseudo_crit}
\end{equation}
Similar definitions are valid for $P^{-}(N)$, and also for $S^{+}(N)$ and $S^{-}(N)$. 
 
 The best estimate of the pseudo-critical point under the event-based ensemble is obtained by taking the ensemble average of 
 the single-instance pseudo-critical points of  Eq.~(\ref{eq:ev_pseudo_threshold1}), thus
 \begin{equation}
    p_c(N) =  \langle p^{+} (N) \rangle \;  .
    \label{eq:ev_pseudo_threshold}
\end{equation}
An analogous definition is obtained by using $p^{-} (N)$ in place of $p^{+} (N)$. The quantity defined in Eq.~(\ref{eq:ev_pseudo_threshold}) still obeys the same scaling relationship as of Eq.~(\ref{eq:pseudo_scaling}).
One can also measure the standard deviation of the single-istance pseudo-critical point, i.e.,
\begin{equation}
    \sigma_{p_c}(N) =  \sqrt{\langle \left[p^{+} (N)\right]^2 \rangle - \left[ \langle p^{+} (N) \rangle \right]^2}\;  .
    \label{eq:ev_std_pseudo_threshold}
\end{equation}
Such a quantity obeys the scaling relation
\begin{equation}
    \sigma_{p_c}(N) \sim N^{-1/\bar{\nu}} \; .
    \label{eq:ev_scaling_std_pseudo_threshold}
\end{equation}

The single-instance pseudo-critical value $P^{+}(N)$ obeys the FSS distribution
 \begin{equation}
    \textrm{Prob.} \left[ P^{+}(N) \right]  = Q^{(1)} \left[  P^{+}(N) \, N^{\beta/\bar\nu} \right] \; .
    \label{eq:ev_pseudo_crit_distr1}
\end{equation}
For the average cluster size, we have instead
\begin{equation}
    \textrm{Prob.} \left[ S^{+}(N) \right]  = Q^{(2)} \left[  S^{+}(N) \, N^{-\gamma/\bar\nu} \right] \; .
    \label{eq:ev_pseudo_crit_distr2}
\end{equation}
Here, $Q^{(1)}(\cdot)$ and $Q^{(2)}(\cdot)$ are universal scaling distributions.
In particular, the average values of such distributions obey the scaling relationships
\begin{equation}
    P_c(N) =  \left\langle P^{+}(N) \right\rangle  \sim N^{-\beta/\bar{\nu}} 
    \label{eq:ev_pseudo_crit_scaling1}
\end{equation}
and
\begin{equation}
    S_c(N) =  \left\langle S^{+}(N) \right\rangle  \sim N^{\gamma/\bar{\nu}} \; ,
    \label{eq:ev_pseudo_crit_scaling2}
\end{equation}
respectively. Analogous scaling relationships are valid also for the standard deviation of the distributions  $Q^{(1)}(\cdot)$ and $Q^{(2)}(\cdot)$, i.e., 
\begin{equation}
    \sigma_{P_c}(N) =  \sqrt{\left\langle \left[P^{+}(N)\right]^2 \right\rangle - \left[\left\langle P^{+}(N)\right\rangle \right]^2} \sim N^{-\beta/\bar{\nu}} 
    \label{eq:ev_pseudo_crit_scaling1a}
\end{equation}
and
\begin{equation}
    \sigma_{S_c}(N) =  \sqrt{\left\langle \left[S^{+}(N)\right]^2 \right\rangle - \left[\left\langle S^{+}(N)\right\rangle \right]^2} \sim N^{\gamma/\bar{\nu}} \; .
    \label{eq:ev_pseudo_crit_scaling2a}
\end{equation}
All the above equations are valid also if we replace $P^{+}(N)$ with $P^{-}(N)$ and $S^{+}(N)$ with $S^{-}(N)$.

\subsection{Critical properties of the shortest-path-percolation in the event-based ensemble}

The following applies regardless if one consider the definition of 
single-instance pseudo-critical point of Eq.~(\ref{eq:ev_pseudo_threshold1}) or of 
Eq.~(\ref{eq:ev_pseudo_threshold2}).
Different estimates of the critical exponents are obtained either by using the scaling of the average of the various observables, or their standard deviations. The numerical results summarized in Tables~\ref{tab:FSS_post_event_full} and~\ref{tab:FSS_pre_event_full} are organized accordingly.

\begin{figure}[!htb]
    \centering
    \includegraphics[width=\columnwidth]{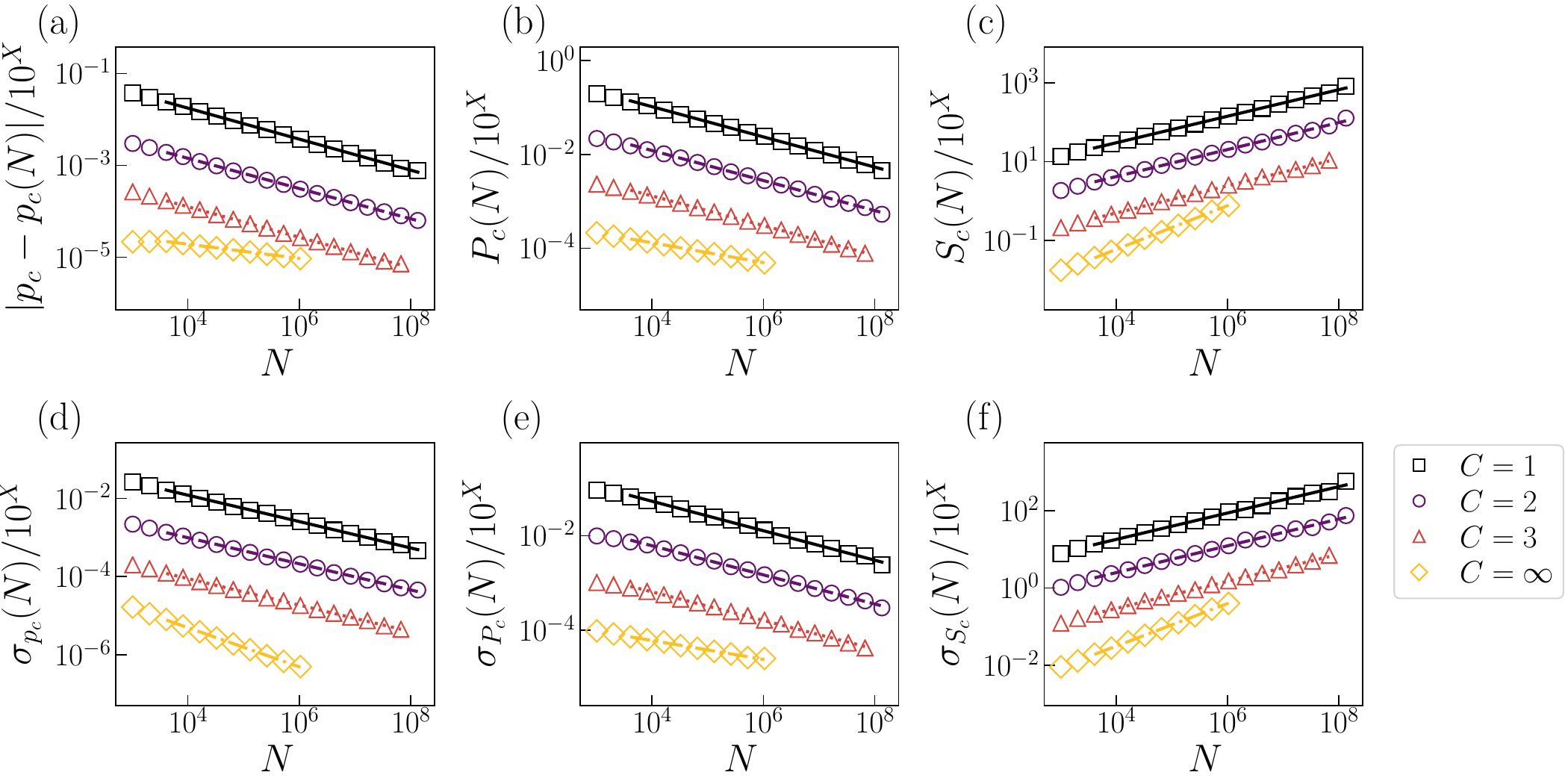}
    \caption{
    Finite-size scaling analysis in the event-based ensemble using the pseudo-critical point of Eq.~(\ref{eq:ev_pseudo_threshold1}).
    Note that an additional factor $10^X$ was used for visual clarity. We use $X=0, 1, 2, 3$ for $C=1, 2, 3, N$, respectively. In each panel, we display as symbols the average value of the metric. Error bars are not displayed 
    to avoid overcrowding the figure.
    Lines appearing in the plot are best fits obtained via simple linear regression 
    on log-transformed variables, see text for details.
    (a) Scaling of Eq.~(\ref{eq:pseudo_scaling}) to estimate $1/\bar\nu$. (b) Scaling at the critical point for the percolation strength, i.e, $P(p=p_c)$ {\it vs} $N$, to estimate $\beta/\bar\nu$. (c) Scaling at the critical point for the average cluster size, i.e, $S(p=p_c)$ {\it vs} $N$, to estimate $\gamma/\bar\nu$. (d) Scaling of Eq.~(\ref{eq:ev_scaling_std_pseudo_threshold}) to estimate $1/\bar\nu$. (e) Scaling at the pseudo-critical point for the percolation strength, i.e., $P(p=p_c(N))$ {\it vs} $N$, to estimate $\beta/\bar\nu$. (f) Scaling at the pseudo-critical point for the average cluster size, i.e., $S(p=p_c(N))$ {\it vs} $N$, to estimate $\gamma/\bar\nu$. Estimated exponents are reported in Table~\ref{tab:FSS_post_event_full}.}
    \label{fig:SM-FSS-post_event}
\end{figure}

\begin{figure}[!htb]
    \centering
    \includegraphics[width=\columnwidth]{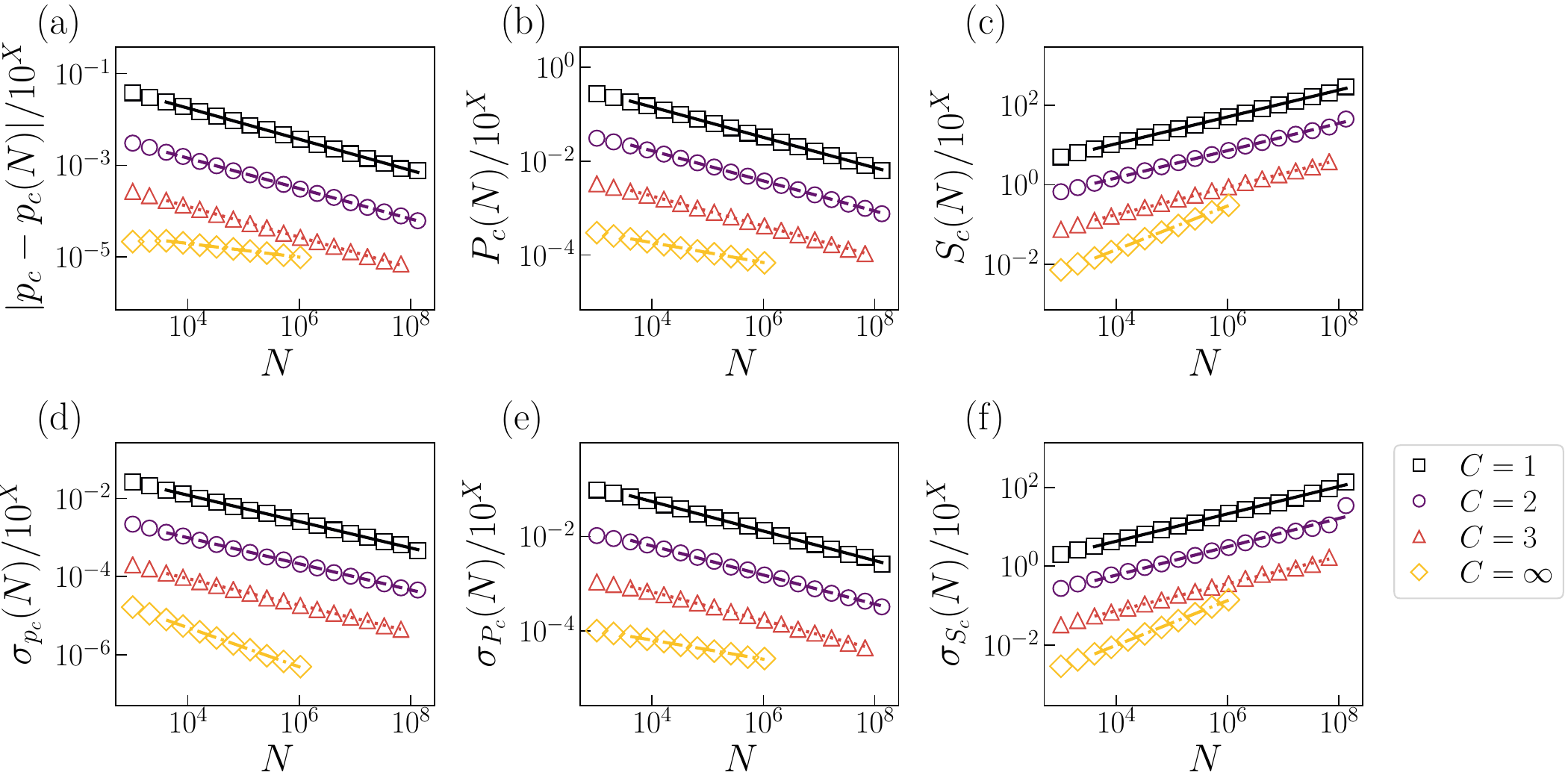}
    \caption{
    Finite-size scaling analysis in the event-based ensemble using the pseudo-critical point of Eq.~(\ref{eq:ev_pseudo_threshold2}).
    Note that an additional factor $10^X$ is used for visual clarity. We use $X=0, 1, 2, 3$ for $C=1, 2, 3, N$, respectively. In each panel, we display as symbols the average value of the metric. Error bars are not displayed 
    to avoid overcrowding the figure. Lines appearing in the plot are best fits obtained via simple linear regression 
    on log-transformed variables, see text for details.
    (a) Scaling of Eq.~(\ref{eq:pseudo_scaling}) to estimate $1/\bar\nu$. (b) Scaling at the critical point for the percolation strength, i.e, $P(p=p_c)$ {\it vs} $N$, to estimate $\beta/\bar\nu$. (c) Scaling at the critical point for the average cluster size, i.e, $S(p=p_c)$ {\it vs} $N$, to estimate $\gamma/\bar\nu$. (d) Scaling of Eq.~(\ref{eq:ev_scaling_std_pseudo_threshold}) to estimate $1/\bar\nu$. (e) Scaling at the pseudo-critical point for the percolation strength, i.e., $P(p=p_c(N))$ {\it vs} $N$ to estimate, $\beta/\bar\nu$. (f) Scaling at the pseudo-critical point for the average cluster size, i.e., $S(p=p_c(N))$ {\it vs} $N$, to estimate $\gamma/\bar\nu$. Estimated exponents are reported in Table~\ref{tab:FSS_pre_event_full}.}
    \label{fig:SM-FSS-pre_event}
\end{figure}

We first determine the best estimate of the critical threshold using Eq.~(\ref{eq:pseudo_scaling}). Here, both $p_c$ and $1/\bar{\nu}$ are treated as free parameters to perform the power-law fit, see \FIG{fig:SM-FSS-post_event} (a) and \FIG{fig:SM-FSS-pre_event} (a). Also, we determine the critical exponent $1/\bar\nu$ using the scaling of Eq.~(\ref{eq:ev_scaling_std_pseudo_threshold}), see \FIG{fig:SM-FSS-post_event} (d) and \FIG{fig:SM-FSS-pre_event} (d).

\begin{figure}[!htb]
    \centering
    \includegraphics[width=0.75\columnwidth]{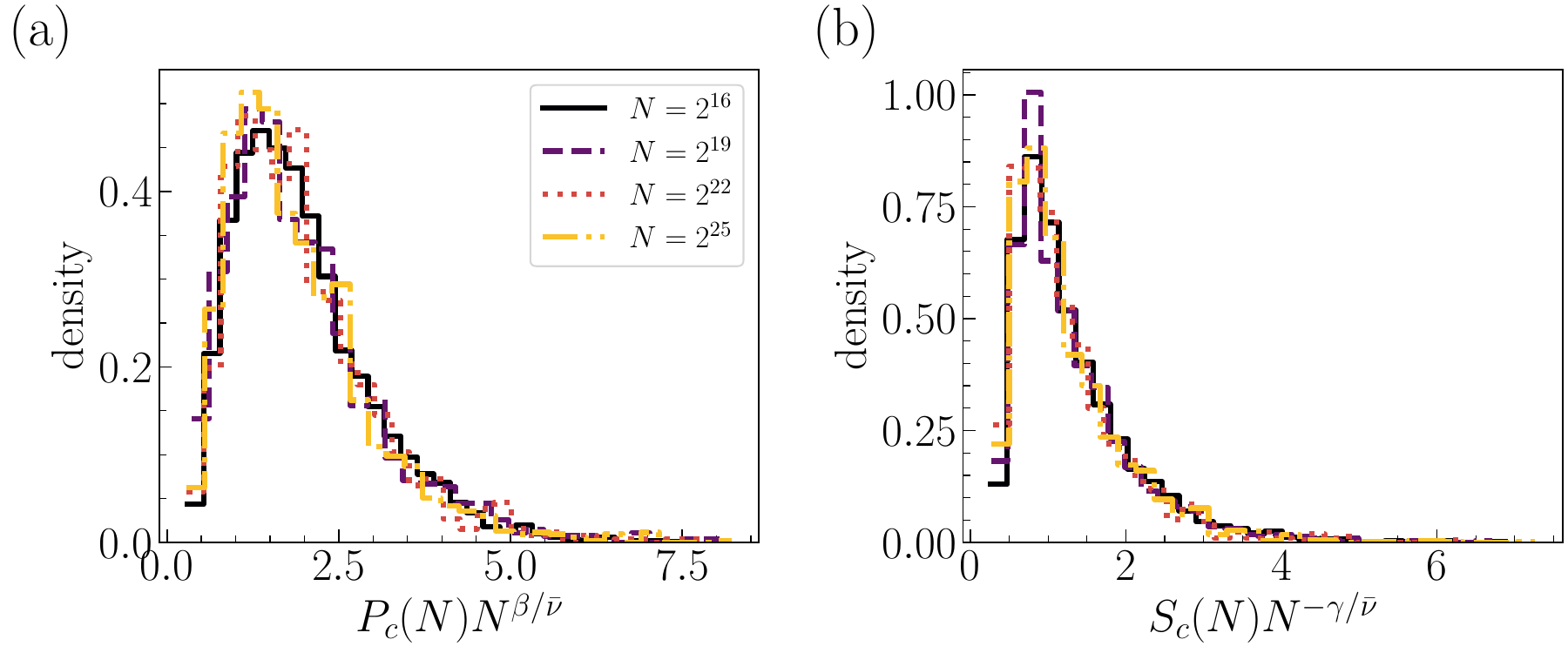}
    \caption{ Shortest-path-percolation transition in the event-based ensemble. Here, we used $C=1$. (a) The distribution of ${P}_{c}(N) \, N^{\beta/\bar{\nu}}$ with $\beta/\bar\nu = 0.32$.
    Different curves correspond to different network sizes $N$. (b) Same as in (a) but for  $S_{c}(N) \,N^{-\gamma/\bar{\nu}}$ with $\gamma/\bar\nu = 0.34$}.
    \label{fig:ev_fss_collapse_C1}
\end{figure}

We then rely on the scaling laws of Eqs.~(\ref{eq:ev_pseudo_crit_scaling1}) and~(\ref{eq:ev_pseudo_crit_scaling2}) to determine the ratio between the critical exponents $\beta/\bar{\nu}$ and $\gamma/\bar\nu$, respectively, see \FIGS{fig:SM-FSS-post_event} (b), (c), and \FIGS{fig:SM-FSS-pre_event} (b), (c). Additional estimates of these exponents are also obtained from  Eqs.~(\ref{eq:ev_pseudo_crit_scaling1a}) and~(\ref{eq:ev_pseudo_crit_scaling2a}), see \FIGS{fig:SM-FSS-post_event} (e), (f), and \FIGS{fig:SM-FSS-pre_event} (e), (f).
Data collapse based on Eqs.~(\ref{eq:ev_pseudo_crit_distr1}) and~(\ref{eq:ev_pseudo_crit_distr2}) is provided in Fig.~3 of the main paper and in \FIG{fig:ev_fss_collapse_C1}.

It's worth mentioning that the two estimates of $1/\bar\nu$ obtained from Eqs.~(\ref{eq:ev_std_pseudo_threshold}) and~(\ref{eq:ev_scaling_std_pseudo_threshold}) are identical for finite $C$, but they differ for $C=\infty$. The latter finding is an agreement with the results of Ref.~\cite{li2023explosive} for explosive percolation, with the caveat that the SPP estimate obtained from Eq.~(\ref{eq:ev_std_pseudo_threshold}) is smaller than the one obtained from Eq.~(\ref{eq:ev_scaling_std_pseudo_threshold}). The exact opposite was instead found for the case of explosive percolation.

\begin{table}[!htb]
\renewcommand{\arraystretch}{1.2}
\centering
\begin{tabular}{cc|ccc|ccc}
 \hline
 \hline
 \multicolumn{2}{c|}{} &  \multicolumn{3}{c|}{average} & \multicolumn{3}{c}{fluctuation}\\
 \hline
 \hline
  $C$ & $p_{c}$ & $1/\bar\nu$ & $\beta/\bar{\nu}$ & $\gamma/\bar{\nu}$ & $1/\bar\nu$ & $\beta/\bar\nu$ & $\gamma/\bar{\nu}$\\ 
 \hline
 1 & {$0.750 \pm 0.001$} & {$0.34 \pm 0.01$} & {$0.32 \pm 0.01$} & {$0.34 \pm 0.01$} & {$0.34\pm 0.01$}& {$0.31\pm 0.01$}& {$0.34\pm 0.01$}\\ 
 2 & {$0.701 \pm 0.001$} & {$0.33 \pm 0.01$} & {$0.32 \pm 0.01$} & {$0.34 \pm 0.01$} & {$0.33\pm0.01$}& {$0.30\pm 0.01$}& {$0.35\pm 0.01$} \\
 3 & {$0.682 \pm 0.001$} & {$0.33 \pm 0.01$} & {$0.32\pm 0.01$} & {$0.34\pm 0.01$ }& {$0.35\pm 0.01$}& {$0.31 \pm 0.01$}& {$0.36\pm 0.01$} \\
 $\infty$ & {$0.646 \pm 0.001$} & {$0.15 \pm 0.01$} & {$0.21 \pm 0.01$} & {$0.55 \pm 0.01$} & {$0.50 \pm 0.01$}& {$0.20\pm 0.01$}& {$0.55\pm 0.01$} \\
 \hline
 \hline 
\end{tabular}
\caption{Critical properties of the shortest-path-percolation (SPP) model in the event-based ensemble. Estimates reported here are obtained via finite-size scaling analysis in the event-based ensemble using  the pseudo-critical point of Eq.~(\ref{eq:ev_pseudo_threshold1}).}
\label{tab:FSS_post_event_full}
\end{table}

\begin{table}[!htb]
\renewcommand{\arraystretch}{1.2}
\centering
\begin{tabular}{cc|ccc|ccc}
 \hline
 \hline
 \multicolumn{2}{c|}{} &  \multicolumn{3}{c|}{average} & \multicolumn{3}{c}{fluctuation}\\
 \hline
 \hline
 $C$ & $p_{c}$ & $1/\bar\nu$ & $\beta/\bar{\nu}$ & $\gamma/\bar{\nu}$ & $1/\bar\nu$ & $\beta/\bar\nu$ & $\gamma/\bar{\nu}$\\ 
 \hline
 1 & {$0.750 \pm 0.001$} & {$0.34 \pm 0.01$} & {$0.32 \pm 0.01$} & {$0.34 \pm 0.01$} & {$0.34\pm0.01$}& {$0.31\pm 0.01$}& {$0.35 \pm 0.01$}\\ 
 2 & {$0.701 \pm 0.001$} & {$0.33 \pm 0.01$} & {$0.32 \pm 0.01$} & {$0.34 \pm 0.01$} & {$0.33 \pm 0.01$}& {$0.30\pm 0.01$}& {$0.35\pm 0.01$} \\
 3 & {$0.682 \pm 0.001$} & {$0.33 \pm 0.01$} & {$0.32 \pm 0.01$} & {$0.34 \pm 0.01$} & {$0.35 \pm 0.01$}& {$0.31\pm 0.01$}& {$0.33\pm 0.01$} \\
 $\infty$ & {$0.646 \pm 0.001$} & {$0.15 \pm 0.01$} & {$0.21 \pm 0.01$} & {$0.55 \pm 0.01$} & {$0.50 \pm 0.01$}& {$0.20 \pm 0.01$}& {$0.57\pm0.01$} \\
 \hline
 \hline 
\end{tabular}
\caption{Critical properties of the shortest-path-percolation (SPP) model in the event-based ensemble. Estimates reported here are obtained via finite-size scaling analysis in the event-based ensemble using the pseudo-critical point of Eq.~(\ref{eq:ev_pseudo_threshold2}).
}
\label{tab:FSS_pre_event_full}
\end{table}

\subsection{Changing control parameter}

\subsubsection{Mapping between control parameters}

\begin{figure}[!htb]
    \centering
    \includegraphics[width=0.75\columnwidth]{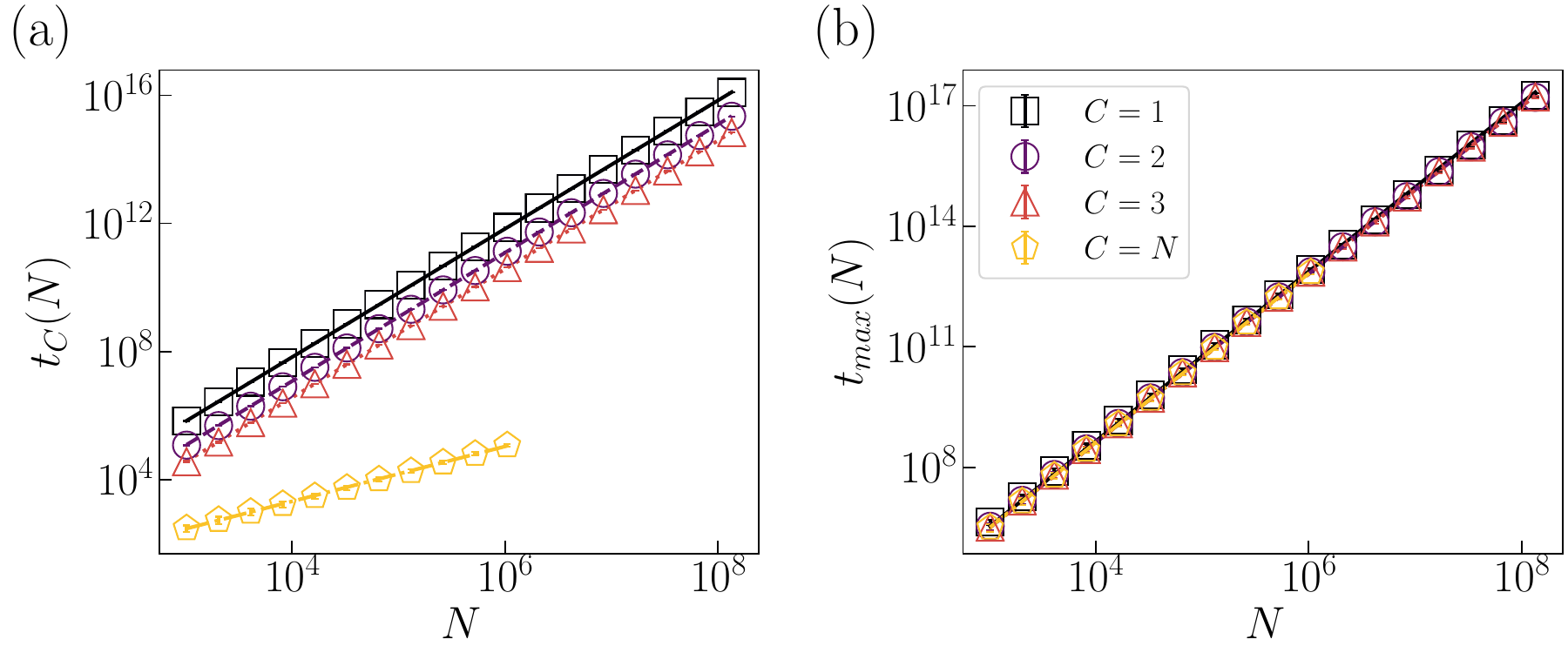}
    \caption{(a) $t_c(N)$ as a function of network size $N$ for different values of $C$. For each $N$ value, we plot the average values of $t_c(N)$ as symbols and the associated standard deviation as error bars. The scaling relation is given by $t_c(N)\sim N^{\theta}$ with $\theta = 2.01 \pm 0.01$ for $C=1$, $\theta = 2.01 \pm 0.01$ for $C=2$, $\theta = 2.01 \pm 0.01$ for $C=3$. For $C=\infty$, we fit $t_c(N)\sim N^{\theta^{\ominus}}$ and obtain $\theta^{\ominus} = 0.86 \pm 0.01$. (b) Maximum value of the raw number of demand $t_{max}(N)$ as a function of the network size $N$ for different values of $C$. 
    The scaling relation is given by $t_{\max}(N)\sim N^{\theta} \, \log(N)$ with $\theta = 2.08 \pm 0.01$ for $C=1$, $\theta = 2.07 \pm 0.01$ for $C=2$, $\theta = 2.06 \pm 0.01$ for $C=3$. For $C=\infty$, we look at $t_{\max}(N) \sim N^{\theta^{\oplus}} \, \log(N)$ and find $\theta^{\oplus} = 2.07 \pm 0.01$}.
    \label{fig:tc_scaling}
\end{figure}

The two natural control parameters of the SPP model, i.e., $p$ and $t$, can be mapped one to the other using Eq.~(\ref{eq:p_t_map}). Typical results of this mapping are displayed in Fig.~4(a) of the main paper. For finite $C$, we find that the map can be written as
\begin{equation}
    p = V \left( t \, N^{-\theta} \right) \; .
    \label{eq:p_t_mpa_finite}
\end{equation}
Here, $V(\cdot)$ is a universal scaling function. An immediate way of estimating the exponent $\theta$ is to look at how $t_c(N)$ scales as function of $N$, see \FIG{fig:tc_scaling} (a). 
Also, we obtain another estimate of $\theta$ from the scaling of
$t_{\max}(N)$, i.e., the total number of random origin-destination pairs that one has to extract in order to remove all edges from an ER network of size $N$, see \FIG{fig:tc_scaling} (b). 

For infinite $C$, we find instead that 
\begin{equation}
p = 
\left\{
\begin{array}{ll}
    V^{-} ( t \, N^{-\theta^{\ominus}} ) & \textrm{ if } t \leq t_c(N)
    \\
    V^{+} ( t \, N^{-\theta^{\oplus}} ) & \textrm{ if } t > t_c(N)
    \end{array}
    \right.
    \; ,
    \label{eq:p_t_mpa_infinite}
\end{equation}
where $V^{-}(\cdot)$ and $V^{+}(\cdot)$ are universal scaling functions valid respectively for the supercritical and subcritical regimes of the SPP transition. 
A good estimate $\theta^{\ominus}$ can be obtained by still looking at how $t_c(N)$ scales with $N$, see \FIG{fig:tc_scaling} (a). We determine the best estimate of $\theta^{\oplus}$ from the scaling of
$t_{\max}(N)$ {\it vs} $N$, see \FIG{fig:tc_scaling} (b). 

\begin{figure}[!htb]
    \centering
    \includegraphics[width=1\columnwidth]{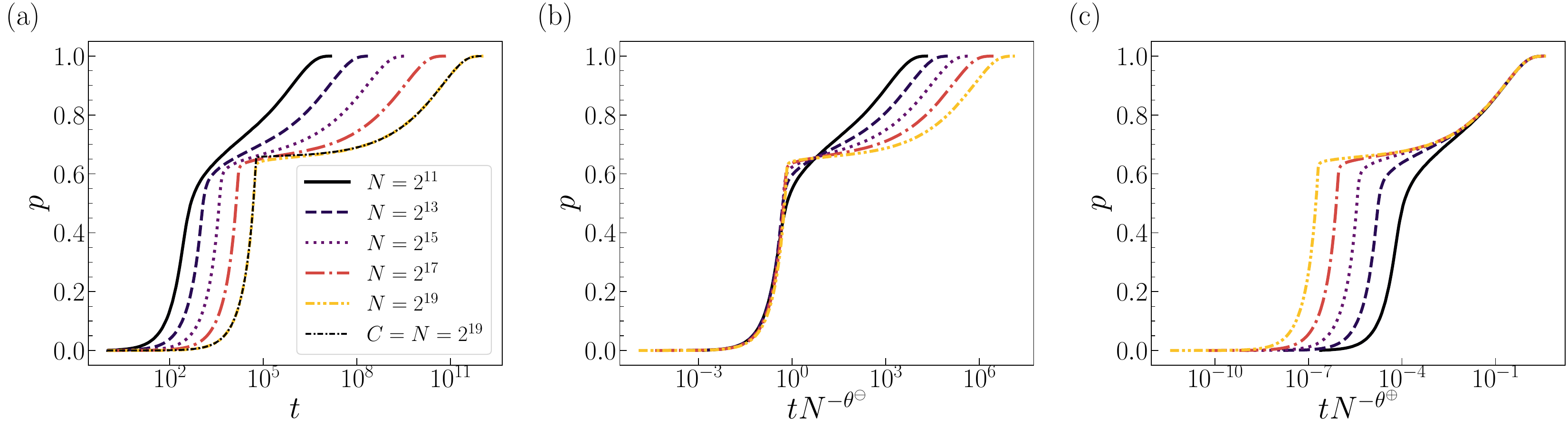}
    \caption{Control parameters in the shortest-path percolation model.
    (a) The fraction of removed edges $p$ is plotted as a function of number of demanding agents $t$ for different values of $N$. Here $C = N^{1/3}$, but we display also the curve for $C=N=2^{19}$. (b) Same as in (a), but the abscissas values are rescaled as $t N^{-\theta^{\ominus}}$, with $\theta^{\ominus} = 0.86$, to obtain a collapse between the various curves. (c) Same as in (a), but the abscissas values are rescaled as $t N^{-\theta^{\oplus}}$, with $\theta^{\oplus} = 2.00$.}
    \label{fig:C03}
\end{figure}

\begin{figure}[!htb]
    \centering
    \includegraphics[width=1\columnwidth]{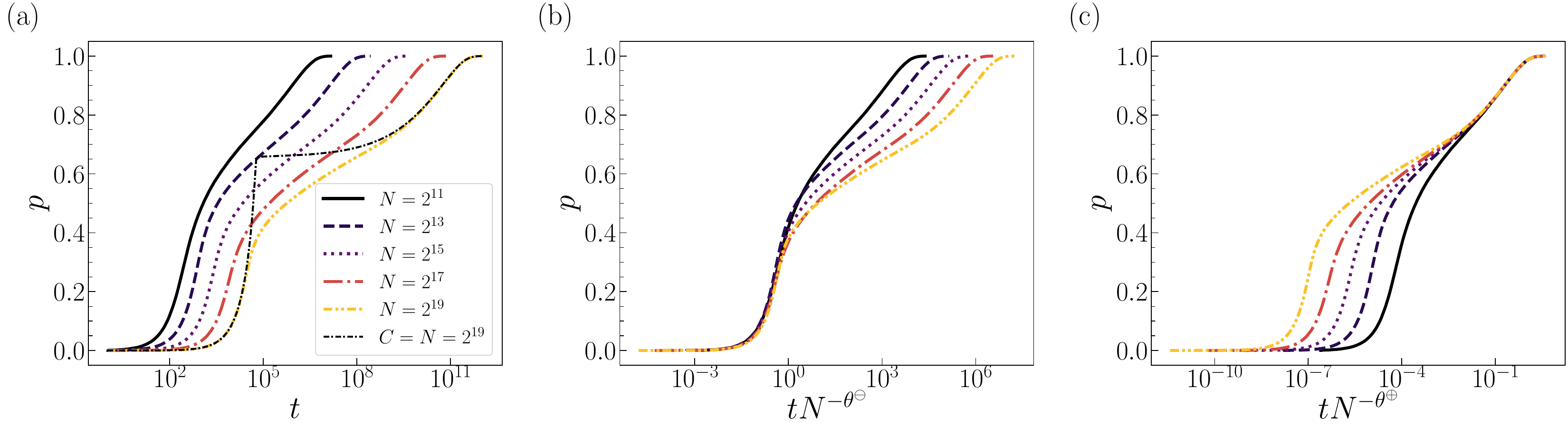}
    \caption{Control parameters in the shortest-path percolation model.
    (a) The fraction of removed edges $p$ is plotted as a function of number of demanding agents $t$ for different values of $N$. Here $C = \log{(N)}$, but we display also the curve for $C=N=2^{19}$. (b) Same as in (a), but the abscissas values are rescaled as $t N^{-\theta^{\ominus}}$, with $\theta^{\ominus} = 0.86$, to obtain a collapse between the various curves. (c) Same as in (a), but the abscissas values are rescaled as $t N^{-\theta^{\oplus}}$, with $\theta^{\oplus} = 2.00$.}
    \label{fig:Clog}
\end{figure}

We test the scaling of Eq.~(\ref{eq:p_t_mpa_infinite}) also for $C = N^{1/3}$ (\FIG{fig:C03}) and $C = \log{(N)}$ (\FIG{fig:Clog}). We do not estimate the critical exponents $\theta^{\ominus}$ and $\theta^{\oplus}$, rather simply verify that curve collapse occur for the same values of the critical exponents obtained for $C=N$.
We see that, for sufficiently large $N$, the curves obtained for $C = N^{1/3}$ and $C=N$ are basically identical.  Differences between the cases $C = \log{(N)}$ and $C=N$ are apparent, however, the corresponding curves are almost identical in both their leftmost and rightmost parts.

\subsubsection{Critical properties of the shortest-path-percolation transition using the control parameter $t$}

All the above theoretical arguments and numerical analyses generalize immediately if one replaces the control parameter $p$ with the control parameter $t$. Specifically, we look at the rescaled control parameter $\tilde{t} = t N^{-2}$ to deal with an intensive quantity similar to $p$. 

Critical threshold values remain finite as along as $C$ is finite.
A vanishing threshold is instead obtained for infinite $C$.

\begin{figure}[!htb]
    \centering
    \includegraphics[width=\columnwidth]{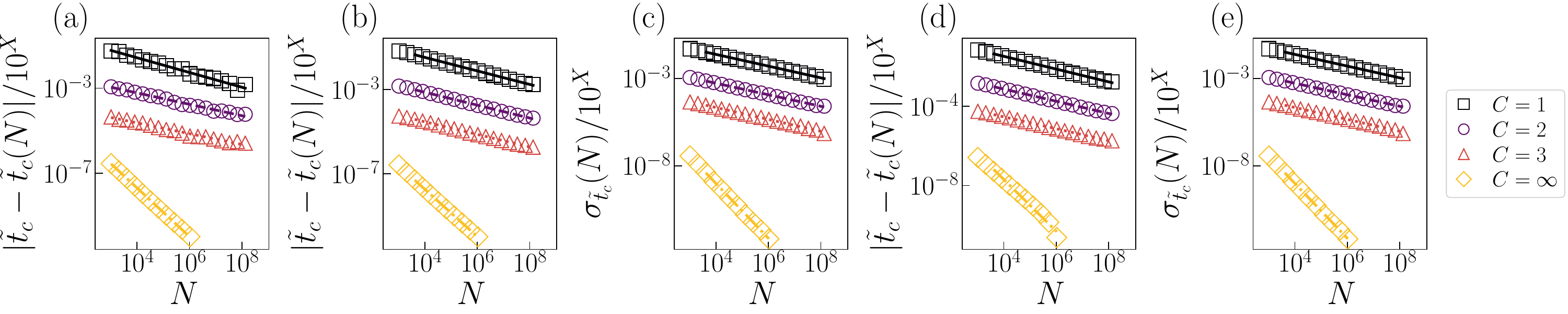}
    \caption{Finite-size scaling analysis with the control parameter $t$. 
    Note that an additional factor $10^X$ is used for visual clarity. We use $X=0, 1, 2, 3$ for $C=1, 2, 3, N$, respectively. 
    In each panel, we display as symbols the average value of the metric. Error bars are not displayed 
    to avoid overcrowding the figure. 
    Lines appearing in the plot are best fits obtained via simple linear regression on log-transformed variables, see text for details.
    (a) Scaling of Eq.~(\ref{eq:pseudo_scaling}) to estimate $1/\bar\nu$ in the conventional ensemble. (b) Similar to (a), but in the event-based ensemble using the pseudo-critical point $t_c (N)$ which corresponds to Eq.~(\ref{eq:ev_pseudo_threshold1}). (c) Scaling of Eq.~(\ref{eq:ev_scaling_std_pseudo_threshold}) using the pseudo-critical point $t_c (N)$ which corresponds to Eq.~(\ref{eq:ev_pseudo_threshold1}). (d) Similar to (b), but using the pseudo-critical point $t_c (N)$ which corresponds to Eq.~(\ref{eq:ev_pseudo_threshold2}). (e) Similar to (c) but using the pseudo-critical point $t_c (N)$ which corresponds to Eq.~(\ref{eq:ev_pseudo_threshold2}).}
    \label{fig:FSS-t-full}
\end{figure}

Due to the one-to-one mapping between $p_c(N)$ and $\tilde{t}_c(N)$,
critical exponent values concerning the order parameter and the average cluster size are unaffected by the specific choice of the control parameter. For finite $C$, some variations are observed for the value of the exponent $1/\bar{\nu}$; these variations are, however, sufficiently small and likely due to finite-size effects. For infinite $C$, we  recover $1/\bar\nu = 2 - \theta^{\ominus}$. Results of the analysis are displayed in \FIG{fig:FSS-t-full} and summarized in Table \ref{tab:FSS_tc_estimation}.

\begin{table}[!htb]
\renewcommand{\arraystretch}{1.2}
\centering
\begin{tabular}{ccc|cc|c|cc|c}
 \hline
 \hline
 \multicolumn{1}{c}{} &  \multicolumn{2}{c|}{conventional} & \multicolumn{2}{c|}{event-based (Eq.~\ref{eq:ev_pseudo_threshold1})} &\multicolumn{1}{c|}{fluctuation}  & \multicolumn{2}{c|}{event-based (Eq.~\ref{eq:ev_pseudo_threshold2})} & \multicolumn{1}{c}{fluctuation}\\
 \hline
 \hline
 $C$ & $\tilde{t}_{c}$ & $1/\bar\nu$ & $\tilde{t}_{c}$ & $1/\bar\nu$ & $1/\bar\nu$ & $\tilde{t}_{c}$ & $1/\bar\nu$ & $1/\bar\nu$\\ 
 \hline
 1 & {$0.693\pm0.001$} & {$0.35\pm0.01$} & {$0.693\pm0.001$} & {$0.32\pm0.01$} & {$0.34\pm0.01$} & {$0.696\pm0.001$} & {$0.32\pm0.01$} & {$0.34\pm0.01$}\\ 
 2 & {$0.125\pm0.001$} & {$0.27\pm0.01$} & {$0.124\pm0.001$} & {$0.31\pm0.01$} & {$0.33\pm0.01$} & {$0.124\pm0.001$} & {$0.31\pm0.01$} & {$0.33\pm0.01$}\\
 3 & {$0.039\pm0.001$} & {$0.25\pm0.01$} & {$0.039\pm0.001$} & {$0.30\pm0.01$} & {$0.34\pm0.01$} & {$0.039\pm0.001$} & {$0.30\pm0.01$} & {$0.34\pm0.01$}\\
 $\infty$ & {$0.000\pm0.000$} & {$1.14\pm0.01$} & {$0.000\pm0.000$} & {$1.11\pm0.01$} & {$1.57\pm0.01$} & {$0.000\pm0.000$} & {$1.31\pm0.01$} & {$1.57\pm 0.01$}\\
 \hline
 \hline 
\end{tabular}
\caption{Estimation of the critical threshold $\tilde{t}_c$ and of the critical exponent $1/\bar\nu$ when $\tilde{t}$ is used as control parameter to monitor the transition. For $C=\infty$, we estimate very small values of the threshold $\tilde{t}_c$, well below the numerical accuracy reported in the table.}
\label{tab:FSS_tc_estimation}
\end{table}




\subsection{Scaling of the size of the $k$-largest cluster}

\begin{figure}[!htb]
    \centering
    \includegraphics[width=0.75\columnwidth]{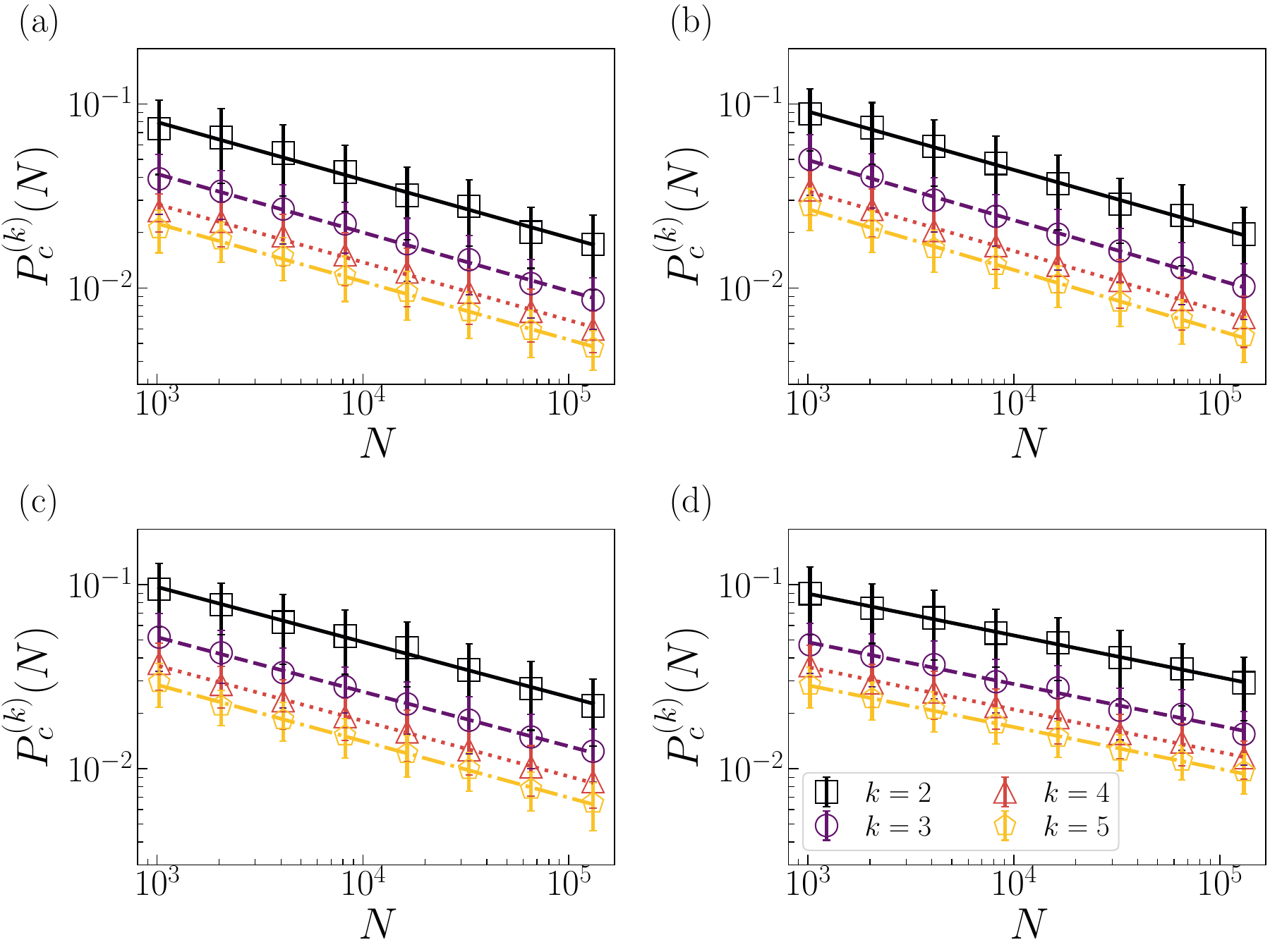}
    \caption{Scaling of the relative size of the $k$-th largest cluster. Results reported here are valid for $k=2, 3, 4$ and $5$. We display as points the average values of the relative size of the $k$-th largest cluster, and with the error bars their standard deviations. Here, we are using 100 samples. The different panels correspond to different values of the model parameter $C$. Specifically, (a) $C=1$, (b) $C=2$, (c) $C=3$, and (d) $C=\infty$. See Table \ref{tab:k_largest_cluster} for detailed information.}
    \label{fig:k_largest_cluster}
\end{figure}

The scaling of Eq.~(\ref{eq:ev_pseudo_crit_scaling1}) valid for the relative size of the largest cluster applies also to the relative size of the $k$-th largest cluster, i.e.,
\begin{equation}
    P^{(k)}_c (N)  = \langle \max_p P^{(k)} (p, N) \rangle \sim N^{-\beta/\bar{\nu}} \; .
    \label{eq:ev_kth_largest_cluster}
\end{equation}
In \FIG{fig:k_largest_cluster}, we show that the scaling holds for $C=1, 2, 3$ and $C = \infty$, for $k=2, 3, 4, $ and $5$. Estimated exponents are also reported in Table~\ref{tab:k_largest_cluster}. As seen in \FIG{fig:k_largest_cluster} and Table~\ref{tab:k_largest_cluster}, we could clearly obtain identical exponents from Eq.~(\ref{eq:ev_kth_largest_cluster}).

\begin{table}[!htb]
\renewcommand{\arraystretch}{1.5}
\centering
\begin{tabular}{ccccc}
 \hline
 \hline
 $C$ & $k=2$ & $k=3$ & $k=4$ & $k=5$\\ 
 \hline
 1 & {$0.31\pm0.01$} & {$0.32\pm0.01$} & {$0.32\pm0.01$} & {$0.32\pm0.01$}  \\ 
 2 & {$0.32\pm0.01$} & {$0.33\pm0.01$} & {$0.32\pm0.01$} & {$0.33\pm0.01$}  \\ 
 3 & {$0.30\pm0.01$} & {$0.30\pm0.01$} & {$0.30\pm0.01$} & {$0.31\pm0.01$} \\ 
 $\infty$ & {$0.23\pm0.01$} & {$0.23\pm0.01$} & {$0.23\pm0.01$} & {$0.23\pm0.01$} \\ 
 \hline
 \hline 
\end{tabular}
\caption{Estimation of the exponent $\beta/\bar\nu$ from \EQ{eq:ev_kth_largest_cluster}.}
\label{tab:k_largest_cluster}
\end{table}


\section{Simulating the Shortest-path-percolation model} \label{apendix:algorithm}

The inputs of the algorithm are: (i) a graph $\mathcal{G}_1 = (\mathcal{V}, \mathcal{E}_1)$ composed of $N = |\mathcal{V}|$ nodes and $E_1= |\mathcal{E}_1|$ edges; (ii) a list of $T$ origin-destination pairs, i.e., $(o_1 \to d_1, o_2 \to d_2, \ldots, o_t \to d_t, \ldots, o_T \to d_T)$; the maximum budget $C$ available to each agent. We assume that the graph $\mathcal{G}_1$ is unweighted and symmetric.

We denote with $\mathcal{G}_{t} = (\mathcal{V}, \mathcal{E}_{t})$ the graph available to the $t$-th agent when the agent demands an itinerary for its origin-destination $o_t \to d_t$. We set $t=1$ and iterate the following:

\begin{enumerate}
    \item We find all shortest paths 
    between the origin $o_t$ and the destination $d_t$; the shortest paths are found on the graph $\mathcal{G}_{t}$ using only edges in the set $\mathcal{E}_{t}$. Denote with $Q_t$ the length of such shortest path. We copy the graph $\mathcal{G}_{t+1} \mapsto \mathcal{G}_{t}$.
    
    \item If there is no path from $o_t$ to $d_t$ or if $Q_t > C$, we proceed to point 4.

    \item If a path from $o_t$ to $d_t$ exists and $Q_t \leq C$, then we select one of the shortest paths between $o_t$ and $d_t$ at random. We denote this path as $(i_0 = o_t, i_1, i_2, \ldots, i_{Q_t-1}, i_{Q_t} = d_t)$,. All edges in the path are removed from the current graph, i.e., $\mathcal{E}_{t+1} \mapsto \mathcal{E}_{t} \setminus \bigcup_{q=1}^{Q_t} (i_{q-1},  i_q)$.

    \item We increase $t \mapsto t+1$ and go back to point 1.
\end{enumerate}

The former algorithm is iterated until $t \leq T$. 

A random shortest path between a generic pair of nodes $o$ and $d$ can be efficiently selected by first applying a breadth-first search (BFS) from $o$ to $d$. 
We perform a BFS with maximum depth $C$,
as only shortest paths of length at most equal to $C$ are considered. 
If node $d$ is not reached 
with a BFS with maximum depth $C$
then no shortest path shorter than $C$ exists between $o$ and $d$. While performing BFS, we construct a directed acyclic graph (DAG) that contains the entire information about all eventual shortest paths between $o$ and $d$. Specifically, for a generic node $i$ in the DAG, we store the information about its parents in the set $\mathcal{P}_i$. Also, we keep track of the number of shortest paths $s_i$ that originate in node $o$ and arrive to node $i$. Once BFS is completed, we sample one of the shortest paths between $o$ and $d$ by performing a random walk from $d$ to $o$ on the DAG.  At each generic node $i$ along the sampled walk, we walk back towards $o$ by randomly selecting a parent of $i$ from the set $\mathcal{P}_i$. The parent $j \in \mathcal{P}_i$ is selected proportionally to the pre-computed number of shortest paths $s_j$ passing thorough it.

\subsection{Shortest-path percolation for randomly selected pairs of nodes}

In our paper, we study the case in which the SPP model assumes that
 a random origin-destination pair is selected at a time.
The naive implementation of this procedure could be computationally inefficient as potentially many of the randomly generated pairs may not satisfy the above requirement, thus leading to a very large number of attempts before a suitable pair is effectively generated. The issue is exacerbated 
for finite $C$ and 
when the graph progressively becomes sparser/disconnected due to the removal of edges. 

To avoid the computational issue and be able to simulate the model in sufficiently large networks, we proceed as follows.
We first construct the set of potentially suitable origin-destination pairs, i.e., $\mathcal{L} = \{o_1 \to d_1, o_2 \to d_2, \ldots, o_z \to d_z, \ldots, o_Z \to d_Z \}$, as the set of all origin-destination nodes at distance less than or equal to $C$ in the graph $\mathcal{G}_1$.
Nodes in the network are labeled from $1$ to $N$, in the set $\mathcal{L}$ we have only pairs  $o_z < d_z$ to avoid double counting. We set $t=1$, and then apply the following:

\begin{enumerate}
     \item We extract a random number $\tau$ from the geometric distribution
     \begin{equation}
     P(\tau) = \frac{2L}{N(N-1)} \, \left[ 1 - \frac{2L}{N(N-1)} \right]^{\tau-1} \; ,
     \label{eq:geometric}
 \end{equation}
 where $L$ is the number of pairs in the set $\mathcal{L}$.
 The probability of success in the geometric distribution is 
 $\frac{2L}{N(N-1)}$, and $\tau$ represents the number of random pairs that one needs to generate in order to find a pair that belongs to $\mathcal{L}$. 
 We create the graph $\mathcal{G}_{t + \tau} \mapsto \mathcal{G}_{t}$. Then, we increase $t \mapsto t + \tau$.

 \item We select a random pair, say $o_t \to d_t$, from the list $\mathcal{L}$. We apply the shortest-path-percolation algorithm by first determining whether $o_t$ and $d_t$ are in the same connected component and finding all eventual shortest paths connecting the two nodes. We then remove all edges along the shortest path between $o_t$ and $d_t$ only if such a shortest path exists and has length smaller or equal than $C$. Similarly to what we described in the preceding section, this step is performed on the graph $\mathcal{G}_{t}$. 
The result of the operation is the graph $\mathcal{G}_t$ containing $E_{t}$ edges.

 \item We verify that the pair $o_t \to d_t$ can still belong to the list $\mathcal{L}$ by measuring the shortest-path distance between $o_t$ and $d_t$ in the graph $\mathcal{G}_t$. If such a distance is larger than $C$, we remove the pair $o_t \to d_t$ from $\mathcal{L}$, i.e., 
 $\mathcal{L} \mapsto \mathcal{L} \setminus \left\{o_t \to d_t\right\}$. If the pair stays in the set, it will be re-considered in next iterations of the algorithm; if the pair is removed from $\mathcal{L}$, then the probability of Eq.~(\ref{eq:geometric}) is automatically updated to account for it at the next stage of the algorithm.

 \item Until the list $\mathcal{L}$ is not empty, we go back to point 1. 

\end{enumerate}

To generate a random number from the distribution of Eq.~(\ref{eq:geometric}), we use 
\[
\tau = 1 + \left\lfloor \log (U) /  \log \left[ 1 - \frac{2L}{N(N-1)} \right]  \right\rfloor \; ,
\]
for $0 < \frac{2L}{N(N-1)} < 1$, and $\tau = 1$ for $\frac{2L}{N(N-1)} = 1$. In the above expression, $U$ is a random variate extracted from the uniform distribution
defined in the range $(0, 1)$, and $
\lfloor \cdot \rfloor$ is the floor function.

We remark that the above procedure is more efficient than the naive implementation of the SPP model only if $Z \leq {N \choose 2}$. This condition is true only when $C \leq \log{ (N/\bar{k})} $ in ER  graphs with size $N$ and average degree $\bar{k}$. Specifically, the computational complexity of the two implementations of the model is identical even for $Z = {N \choose 2}$, however, the proposed alternative implementation is subject to a greater space complexity since it relies on the fact that a list  composed of $Z$ elements is stored in memory.

For finite $C$, we use this efficient strategy from the beginning of the simulation. We have in fact that $L \simeq N \bar{k}^C / 2$, so that space complexity scales linearly with $N$; we find that the amount of time required to simulate the SPP model scales slightly super-linearly with $N$, see Figure~\ref{fig:alg_complexity}.

\subsection*{Mixed strategy for the simulation of the infinite-$C$ SPP model}

For the infinite-$C$ SPP model, we adopt the following mixed strategy:

\begin{itemize}
    
    \item[A)] Starting from the input network, we run the naive implementation of the SPP algorithm until no edges are deleted for $T_*$ consecutive randomly selected pairs of nodes.  

    \item[B)]  We keep removing edges using the alternative version of the algorithm described above. This second part is started from the network remaining after the deletion of the edges occurred at point A.

\end{itemize}

The output is given by the concatenation of the two lists of edges identified  at points A and B, respectively. 

The value of the parameter $T_*$ that determines the condition to pass from point A to point B is chosen in a way that optimizes the trade-off between time and space complexity. If, at a certain stage of the simulation, $\mathcal{L}$ is the set of pairs with distance shorter or equal than $C$, then the probability of extracting less than $T_*$ consecutive random pairs of nodes outside this set is given by the  cumulative distribution of Eq.~(\ref{eq:geometric}), thus
\begin{equation}
    P(\tau < T_*) = \left[ 1 - \frac{2L}{N(N-1)} \right]^{T_*} \; ,
    \label{eq:cum_geometric}
\end{equation}
where we indicated with $L=|\mathcal{L}|$ the size of the set $\mathcal{L}$. We remind that $L$ does not increase during the course of a simulation. Ideally, we would like to set $T_*$ such that we pass from point A to point B when $L \ll N^2$; this ensures that we are able to take advantage of the computational speed up without incurring the computational burden of pre-computing and storing the distance between a large number of pairs of nodes.

For infinite $C$, $L \sim N^2$ from the beginning of the simulation until the critical point is reached. At criticality, we have instead $L \sim N^{2 - 2\beta/\bar{\nu}} + N^{1+\gamma/\bar\nu} \ll N^2$, where $\beta$, $\gamma$ and $\bar\nu$ are the critical exponents associated with the size of the largest cluster and the average cluster size. This follows from the fact that the size of the largest cluster scales as $N^{1 - \beta/\bar{\nu}}$at criticality [Eq.~(\ref{eq:perc_str_critical})], and all the pairs of nodes within such a cluster are part of the set $\mathcal{L}$. Other elements of the set $\mathcal{L}$ are pairs of nodes in finite clusters, which is proportional to the average cluster size $S N \sim N^{1+\gamma/\bar\nu}$,  see Eq.~(\ref{eq:av_cluster}). We note that $2 - 2\beta/\bar{\nu} = 1+\gamma/\bar\nu$, we can therefore write
\begin{equation}
    P(\tau < T_*) \simeq \left[ 1 - N^{-2\beta/\bar\nu} \right]^{T_*} \simeq e^{ - T_*  \, N^{2\beta/\bar\nu}} \; ,
    \label{eq:cum_geometric_approx_b}
\end{equation}
where we used the approximation $\log (1-x) \simeq -x$ for $x \ll 1$.
A reasonable way of stating that $P(\tau < T_*)$ is small enough is to impose $P(\tau < T_*) < 1/V$, with $V$ number of simulations performed, leading to
\begin{equation}
    T_* >  \log{(V)} \, N^{2\beta/\bar\nu} \; .
    \label{eq:minT_infinite}
\end{equation}

\begin{figure}[!htb]
    \centering
    \includegraphics[width=0.8\columnwidth]{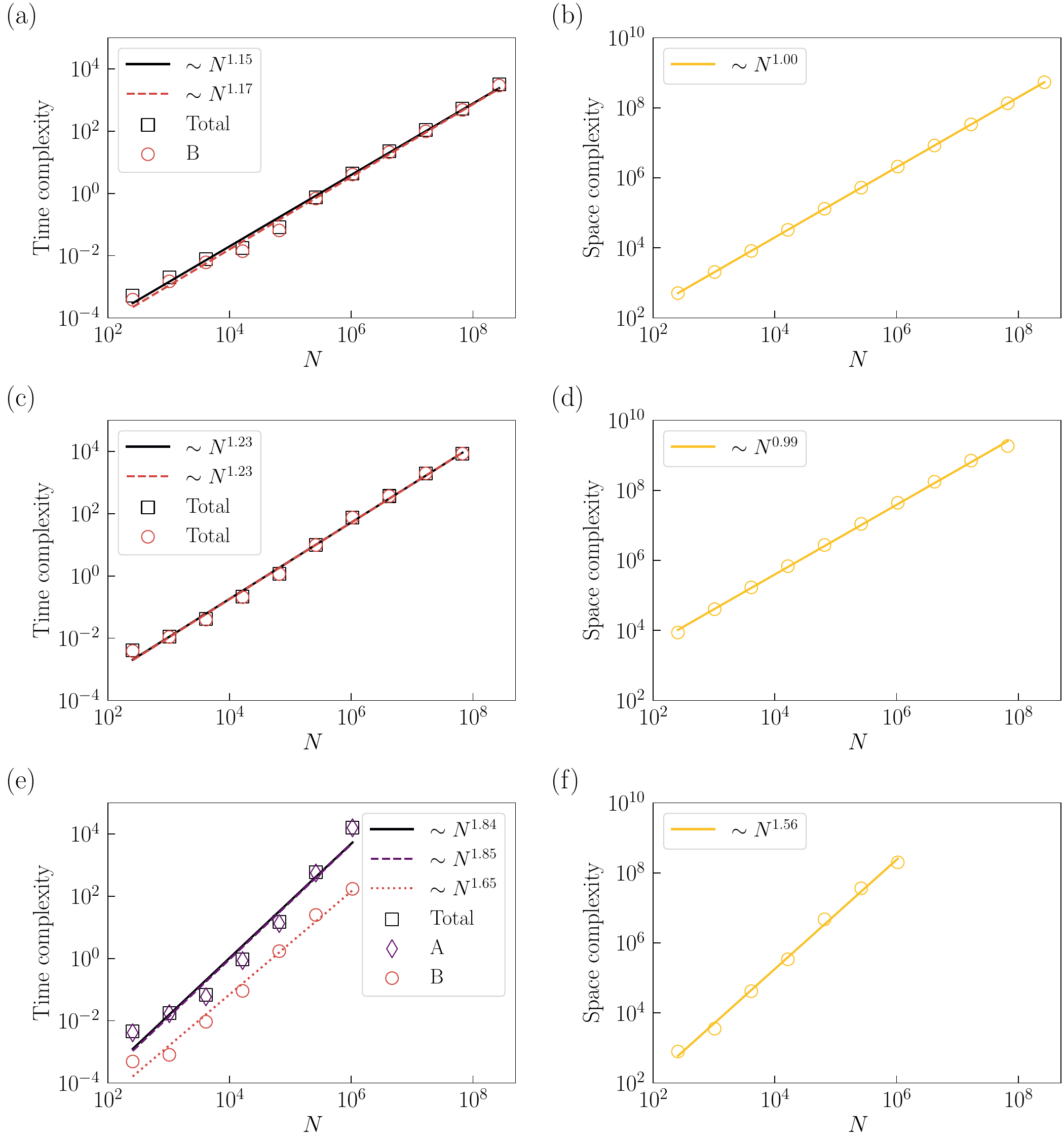}
    \caption{
    Algorithmic complexity of the shortest-path-percolation (SPP) model.
    (a) We plot the amount of time necessary to run a single simulation of the SPP model as a function of the network size $N$. Here, $C=1$. We plot separately the time required for point B of the mixed simulation strategy, and the total time required to draw the entire phase diagram. The total time includes all parts of the algorithm required to draw the phase diagram, including the generation of the graph and the Newman-Ziff portion that allows us to compute the percolation strength and the average cluster size. Data points are fitted with power laws to determine the time complexity of the algorithm. Simulations were performed on a Intel(R) Xeon(R) CPU E5-2690 v4 @ 2.60 GHz and time is measured in seconds. (b) We visualize the size (i.e., number of elements) of the set $\mathcal{L}$ used in part B) of the mixed simulation strategy as a function of the network size for the same set of simulations as in panel (a). Data points are fitted with power laws to determine the space complexity of the algorithm. (c) and (d) are the same as panels (a) and (b), respectively, but for $C=3$. (e) Same as in panels (a) and (c), but for $C=N$. Here, we report also the amount of time required to run part A) of the mixed simulation strategy. (f) Space complexity for the same set of simulations as in (e).}
    \label{fig:alg_complexity}
\end{figure}

We set $T_* = 7 \, N^{0.42}$ corresponding to $V = 1,000$ (i.e., roughly equal to the number of our experiments for most of the $N$ values considered, see Table~\ref{tab:SPP_summary}) and $2\beta/\bar\nu = 0.42$ as measured in our numerical tests, see Table~\ref{tab:summary}.
Before the critical point, the application of point A requires a time that scales as $N^{1 + \theta^{\ominus}} \simeq N^{1.86}$, as we need to select $t_c(N) \sim N^{\theta^{\ominus}}$ random pairs of nodes to reach the critical point, and for each of them we need to find the shortest paths in a time that scales as $N$, see Eq.~(\ref{eq:p_t_mpa_infinite}) and Fig.~\ref{fig:tc_scaling}. Once the critical point is reached, we need to determine the set  $\mathcal{L}$; the operation requires time and space complexity both growing as $N^{2 - 2\beta/\bar{\nu}} \simeq N^{1.58}$, see for example Fig.~\ref{fig:SM-FSS-convention}. The above predictions about time and space complexity of the mixed strategy used to simulate the infinite-$C$ SPP model are validated in Fig.~\ref{fig:alg_complexity}.

\end{document}